\documentclass[12pt]{article}

\pdfoutput=1

\usepackage{scicite}

\usepackage{times}

\usepackage{graphicx,pdfpages,url,pdflscape,caption}

\newenvironment{sciabstract}{%
\begin{quote} \bf}
{\end{quote}}

\newcounter{lastnote}
\newenvironment{scilastnote}{%
\setcounter{lastnote}{\value{enumiv}}%
\addtocounter{lastnote}{+1}%
\begin{list}%
{\setlength{\leftmargin}{.22in}}
{\setlength{\labelsep}{.5em}}}
{\end{list}}

\title{Distances, Luminosities, and Temperatures of the Coldest Known Substellar Objects}

\author
{Trent J.\ Dupuy$^{1,2\ast}$ and Adam L.\ Kraus$^{1,3}$ \\
\\
\normalsize{$^{1}$Harvard-Smithsonian Center for Astrophysics, 60 Garden St, Cambridge, MA 02138, USA}\\
\normalsize{$^{2}$Hubble Fellow}\\
\normalsize{$^{3}$University of Texas at Austin, Astronomy Department, Austin, TX 78712, USA}\\
\\
\normalsize{$^\ast$To whom correspondence should be addressed; E-mail:  tdupuy@cfa.harvard.edu.}
}

\date{}

\newcommand{\Spitzer}{{\sl Spitzer}}

\newcommand{\WISE}{{\sl WISE}}

\newcommand{\Lsun}{\mbox{$L_{\odot}$}}
\newcommand{\Rsun}{\mbox{$R_{\odot}$}}
\newcommand{\Mjup}{\mbox{$M_{\rm Jup}$}}
\newcommand{\Rjup}{\mbox{$R_{\rm Jup}$}}

\newcommand{\arcsec}{\mbox{$^{\prime\prime}$}}

\newcommand{\kms}{\mbox{km\,s$^{-1}$}}
\newcommand{\masyr}{\hbox{mas\,year$^{-1}$}}

\newcommand{\Lbol}{\mbox{$L_{\rm bol}$}}
\newcommand{\Rstar}{\mbox{$R_{\star}$}}

\newcommand{\Teff}{\mbox{$T_{\rm eff}$}}
\newcommand{\logg}{\mbox{$\log{g}$}}
\newcommand{\Vtan}{\mbox{$V_{\rm tan}$}}

\newcommand{\apj}{ApJ}
\newcommand{\apjl}{ApJL}
\newcommand{\mnras}{MNRAS}
\newcommand{\aap}{A\&A}

\newcommand{\apjs}{ApJS}
\newcommand{\aj}{AJ}
\newcommand{\araa}{ARA\&A}
\newcommand{\nat}{Nature}
\newcommand{\pasp}{PASP}

\def\lesssim{\mathrel{\hbox{\rlap{\hbox{\lower4pt\hbox{$\sim$}}}{\raise2pt\hbox{$<$}}}}}
\def\gtrsim{\mathrel{\hbox{\rlap{\hbox{\lower4pt\hbox{$\sim$}}}{\raise2pt\hbox{$>$}}}}}

\def\fdg{\hbox{$.\!\!^\circ$}}
\def\farcm{\hbox{$.\mkern-4mu^\prime$}}
\def\farcs{\hbox{$.\!\!^{\prime\prime}$}}

\begin{document} 


\baselineskip24pt


\maketitle 


\begin{sciabstract}

  The coolest known brown dwarfs are our best analogs to extrasolar
  gas-giant planets. The prolific detections of such cold substellar
  objects in the past two years has spurred intensive followup, but
  the lack of accurate distances is a key gap in our understanding.
  We present a large sample of precise distances based on homogeneous
  mid-infrared astrometry that robustly establish absolute fluxes,
  luminosities, and temperatures. The coolest brown dwarfs have
  temperatures of 400--450\,K and masses $\approx$5--20$\times$ that
  of Jupiter, showing they bridge the gap between hotter brown dwarfs
  and gas-giant planets. At these extremes, spectral energy
  distributions no longer follow a simple correspondence with
  temperature, suggesting an increasing role of other physical
  parameters such as surface gravity, vertical mixing, clouds, and
  metallicity.

\end{sciabstract}


One major goal in astrophysics is to extend previous successes in the
characterization and modeling of stellar atmospheres to the much
cooler atmospheres of extrasolar planets.  A key pathway is the
identification of free-floating objects that not only share common
temperatures with exoplanets but that also share common masses and
thus surface gravities.  In recent years, searches for ever colder
free-floating brown dwarfs---objects with masses below the
hydrogen-fusing mass limit---have steadily pushed the census of the
solar neighborhood to ever lower masses and finally perhaps into the
planetary-mass regime ($\lesssim$13 Jupiter masses).

The detection of large samples of brown dwarfs at the beginning of the
last decade ushered in two, now widely accepted, spectral types
denoted by the letters ``L'' and ``T'' that extend the canonical
OBAFGKM scheme for classifying stars that had stood untouched for
nearly a century.  Over the last two years, candidates for a ``Y''
spectral class have been uncovered in binary surveys
\cite{2011ApJ...730L...9L,2011ApJ...740..108L} and in all-sky imaging
data from \WISE, the {\sl Wide-field Infrared Survey Explorer}
\cite{2011ApJ...743...50C}.  The primary criterion adopted to trigger
this class has been the appearance of ammonia (NH$_3$) absorption in
near-infrared (1--2.5\,$\mu$m) spectra.

Y~dwarfs probe colder atmospheric physics than before, with putative
effective temperatures as low as $\Teff \sim 300$\,K and masses of
$\approx$5--20 Jupiter masses \cite{2011ApJ...743...50C}. If found
orbiting a star, a Y~dwarf would likely be considered a gas-giant
planet.  However, these estimated properties of Y~dwarfs are
speculative given the uncertainty in their temperatures, ages, and
luminosities. Temperatures have only been estimated from model
atmospheres that use incomplete molecular line lists and simple
prescriptions for complex processes like nonequilibrium chemistry and
condensate formation.

An independent approach for determining temperatures is to combine
bolometric luminosities (\Lbol) with evolutionary model-predicted
radii (\Rstar) and apply the Stefan--Boltzmann Law, $\Teff \equiv (4
\pi \sigma \Rstar^2 / \Lbol)^{-1/4}$.  Recent observations of
transiting substellar objects generally support evolutionary model
radius predictions over a wide range of masses
\cite{2008A&A...491..889D,2009Natur.460.1098H,2011ApJ...726L..19A}.
Although many may not be ideal test cases, since they may have formed
via core accretion or have been intensely irradiated, variations in
radii are expected to be relatively small and not strongly influence
our resulting temperatures given the weak dependence on radius ($\Teff
\propto \Rstar^{-1/2}$).  Therefore, the key measurements needed to
determine temperatures via luminosity are accurate distances to
Y~dwarfs, along with a method for computing \Lbol\ from
multi-wavelength photometry.

Trigonometric parallaxes provide the only direct means of measuring
distances to stars.  A star's distance is inversely proportional to
the amplitude of its apparent periodic motion on the sky relative to
more distant background stars, which is due to the Earth's orbital
motion around the Sun.  The amplitude of this effect is small,
0.1~arcseconds for a star at 10~parsec, and thus measuring parallaxes
requires long-term, precise position measurements. We have been using
the Infrared Array Camera (IRAC) on board the \textsl{Spitzer Space
  Telescope} to obtain such astrometry of late-T and Y~dwarfs from
2011--2012.

\Spitzer\ currently trails the Earth by $\approx$2 months in its solar
orbit, and keeping its solar shield directed at the Sun forces the
telescope to observe stars near parallax maximum. By maintaining a
cold temperature \Spitzer\ can obtain sensitive images in the thermal
mid-infrared, where Y~dwarfs emit most of their flux, giving it an
advantage over ground-based near-infrared observations of Y~dwarfs.
We also use an improved correction for the nonlinear optical
distortion of \Spitzer/IRAC that enables $\sim$10$\times$ smaller
residual errors than the correction used by the standard data
pipeline, allowing us to unlock the precision astrometric capabilities
of \Spitzer.

\begin{figure}

\centerline{
\includegraphics[width=2.4in,angle=0]{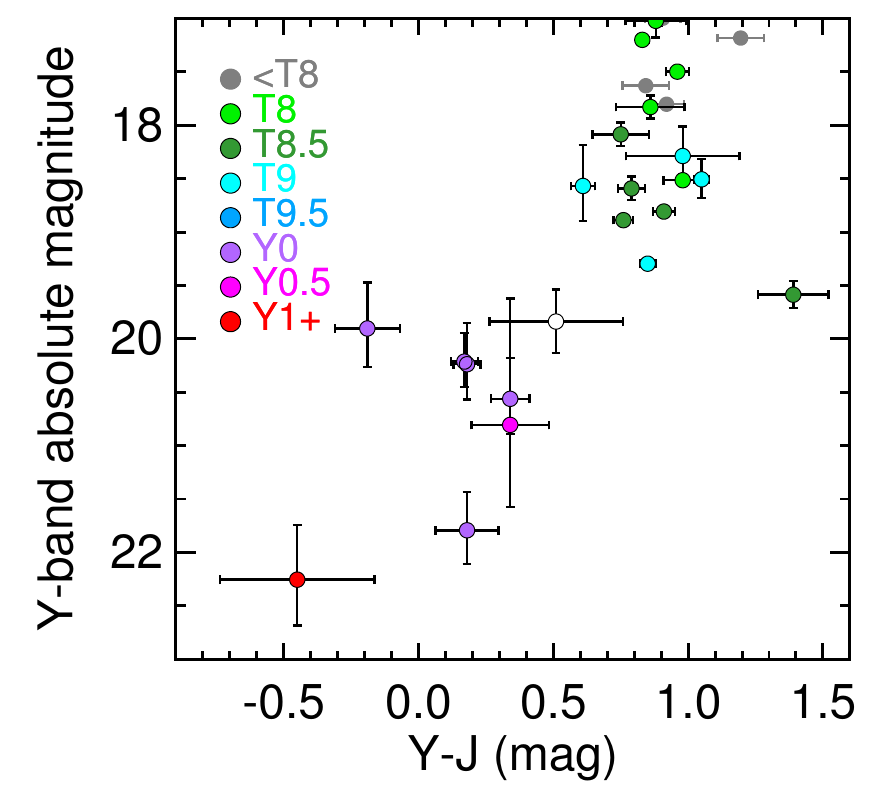}
\hskip 0.0in
\includegraphics[width=2.4in,angle=0]{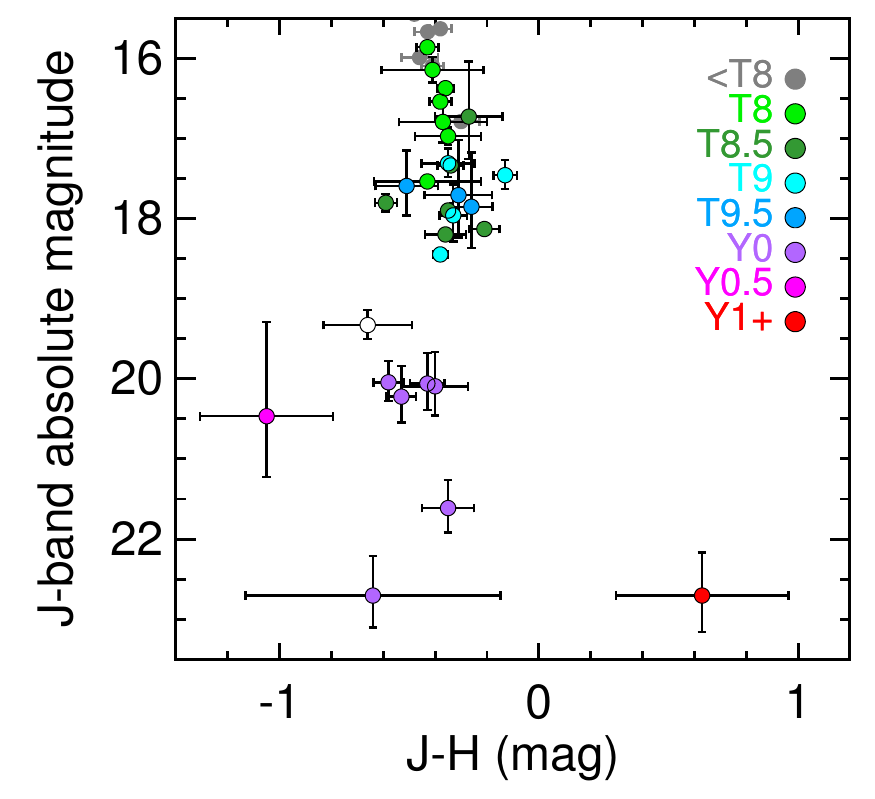}
}
\vskip 0.1in
\centerline{
\includegraphics[width=2.4in,angle=0]{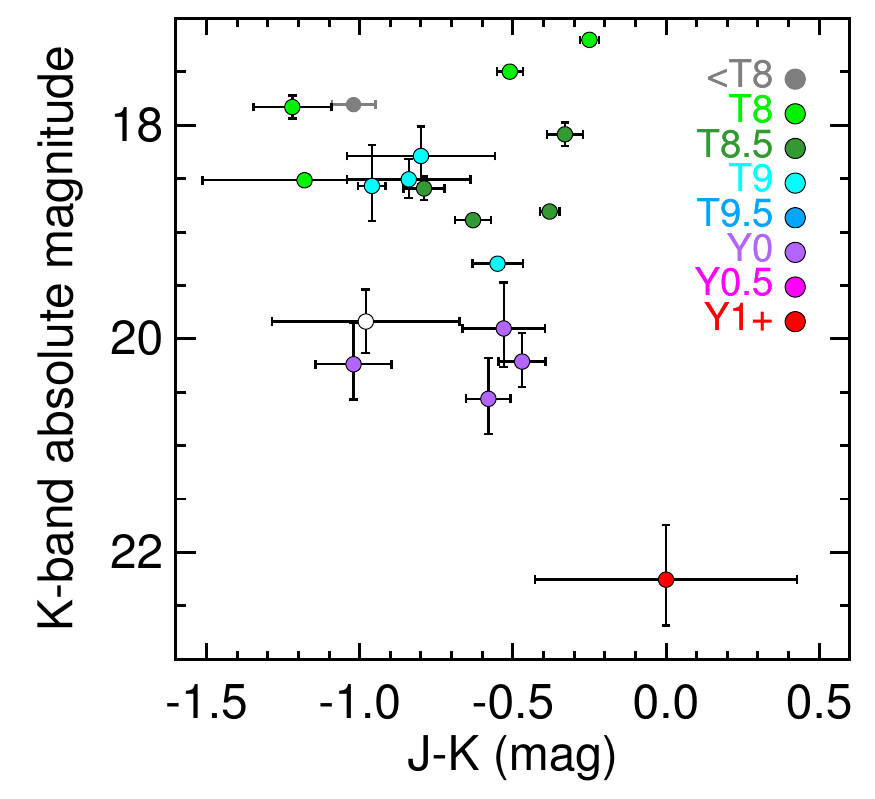}
\hskip 0.0in
\includegraphics[width=2.4in,angle=0]{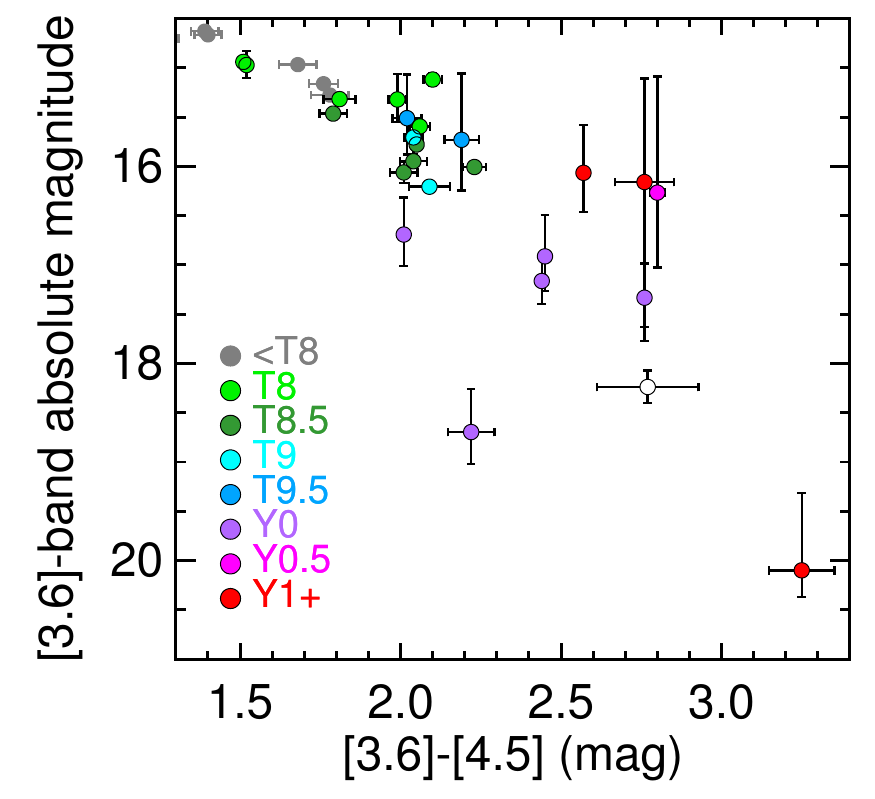}
}

\caption{\normalsize Color--magnitude diagrams for all objects with
  spectral types T8 and later that have direct distance measurements.
  Data points are color coded according to spectral type, with
  open/white points indicating that no spectra are available.  Small
  gray points are earlier type field brown dwarfs. Near-infrared
  photometry is on the Mauna Kea Observatory (MKO)
  system. \label{fig:cmd}}

\end{figure}

\begin{figure}

\centerline{
\includegraphics[width=1.6in,angle=0]{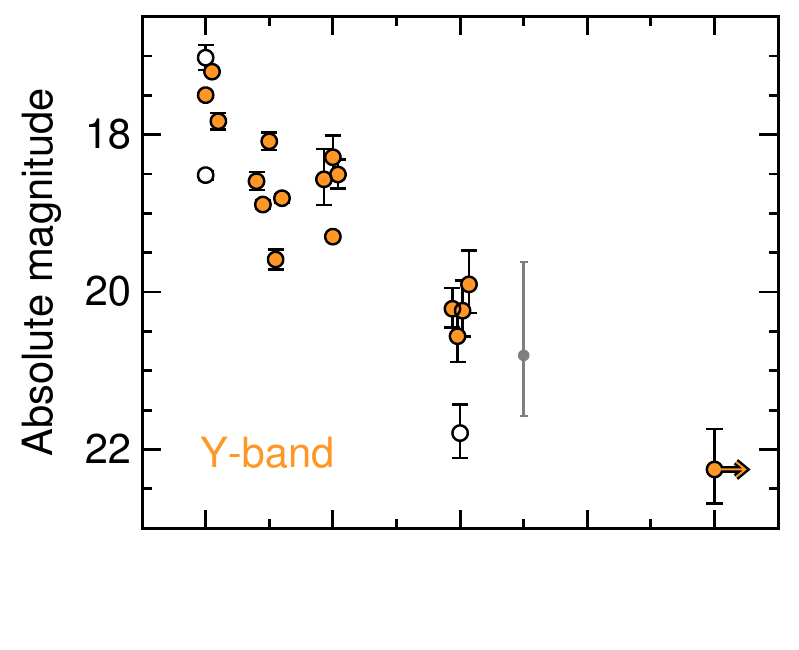}
\includegraphics[width=1.6in,angle=0]{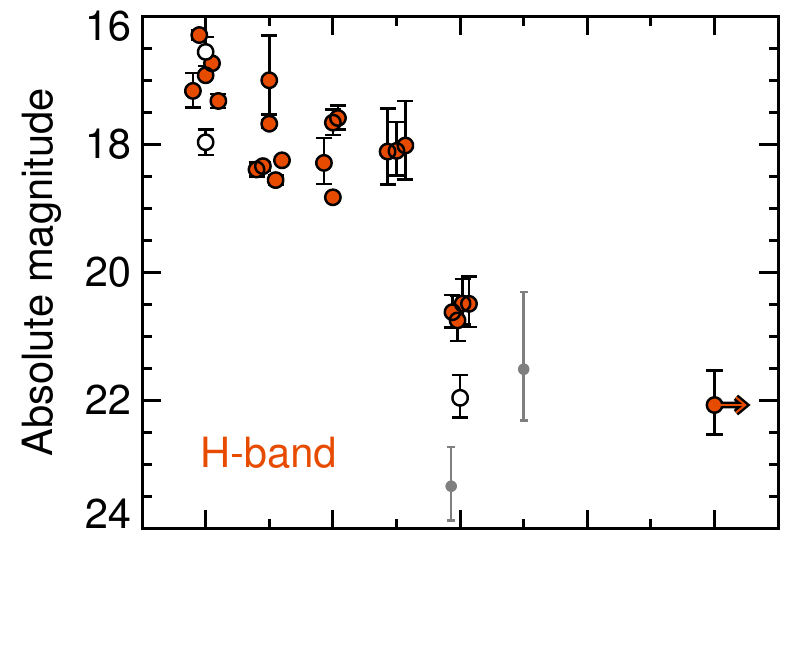}
\includegraphics[width=1.6in,angle=0]{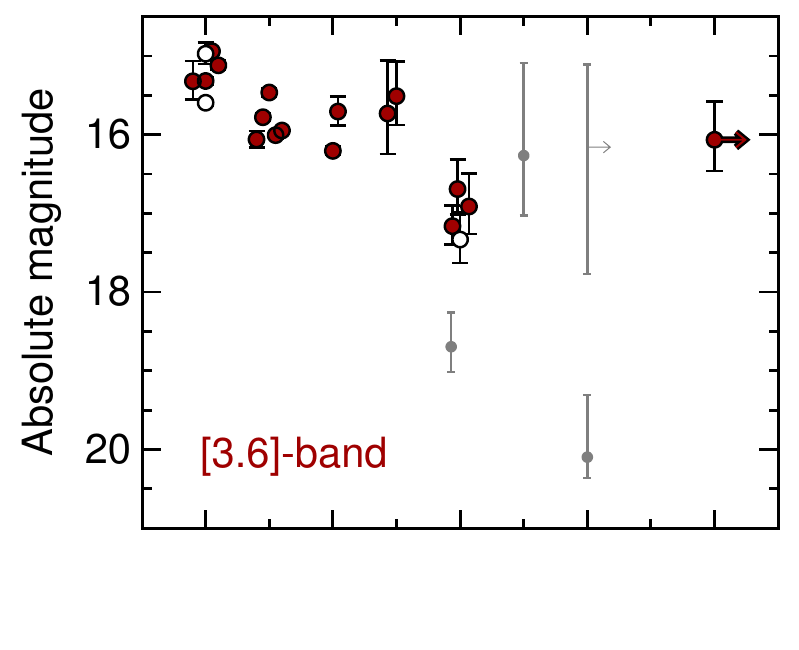}
}
\vskip -0.20in
\centerline{
\includegraphics[width=1.6in,angle=0]{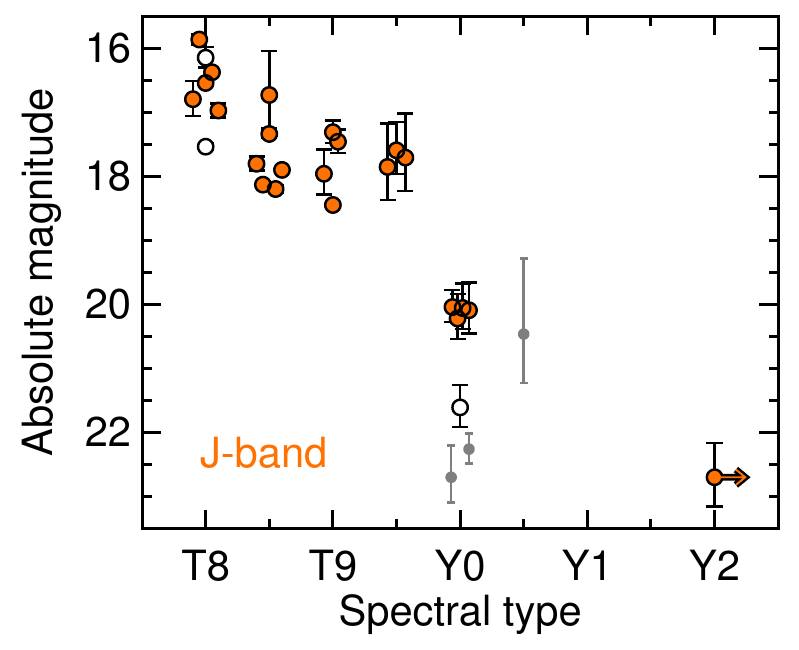}
\includegraphics[width=1.6in,angle=0]{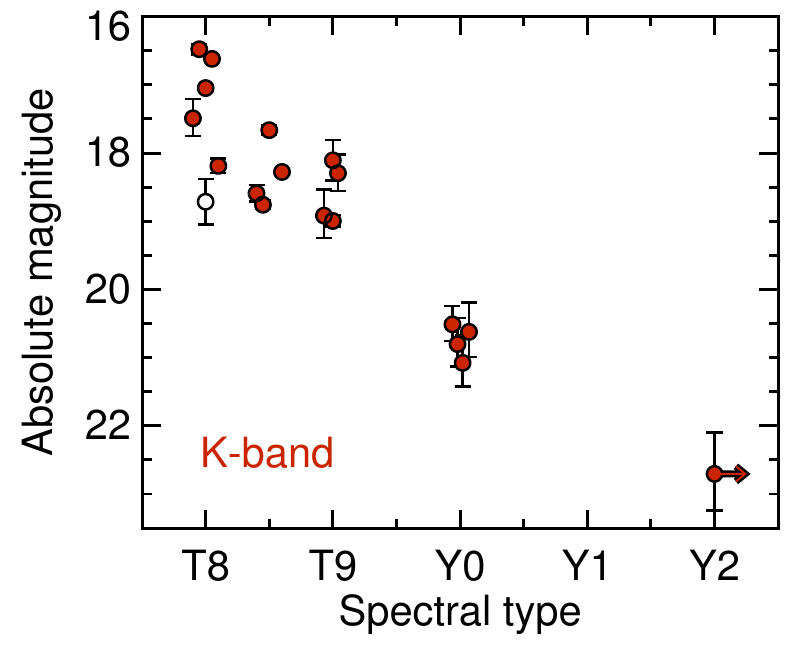}
\includegraphics[width=1.6in,angle=0]{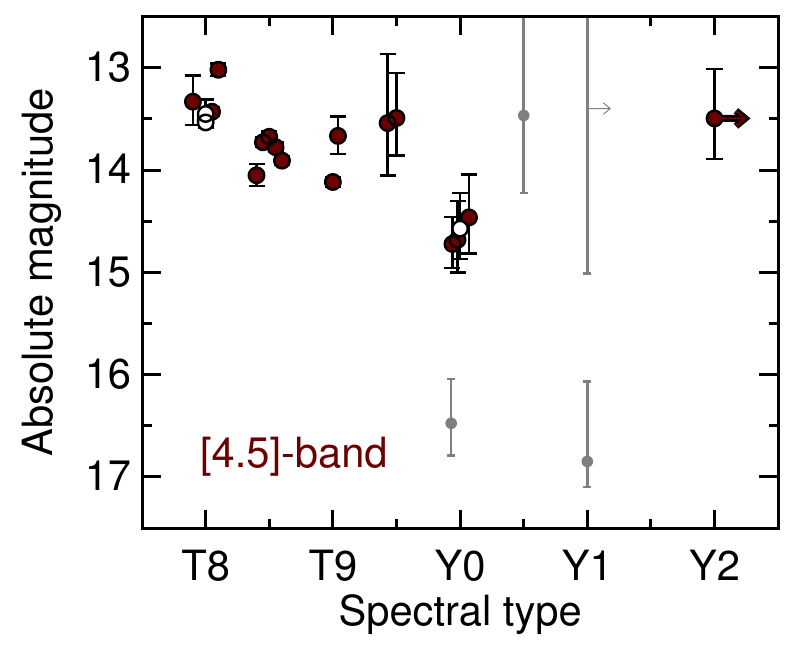}
}

\caption{\normalsize Absolute magnitude as a function of spectral type
  for near-infrared and mid-infrared bandpasses.  Objects typed as
  peculiar are shown as open white symbols. Objects with very
  uncertain distances are plotted with smaller gray symbols. Error
  bars for spectral types are not plotted, and small $x$-axis offsets
  have been added to the spectral types for
  clarity. \label{fig:spt-abs}}

\end{figure}

By combining our parallaxes (Table~S1, Fig.~S\ref{fig:plx}) with
photometry from the literature
\cite{2011ApJS..197...19K,2012ApJ...753..156K,2013ApJ...763..130L}, we
have determined absolute magnitudes in the near-infrared $YJHK$ bands
($\approx$1.0--2.4\,$\mu$m) and \Spitzer's mid-infrared bands at
3.6\,$\mu$m and 4.5\,$\mu$m (Table~S2, Fig.~\ref{fig:cmd},
Fig.~\ref{fig:spt-abs}). For each spectral type bin, we computed the
weighted mean absolute magnitude as well as upper/lower limits on the
amount of intrinsic scatter in the magnitudes (Table~S3).

Objects classified as normal Y0~dwarfs are $\approx$2 magnitudes
($\approx$6$\times$) fainter in the near-infrared compared to the
latest type T~dwarfs, yet they generally share very similar colors.
The most notable exception is that the $Y-J$ colors become much bluer
for Y~dwarfs \cite{2013ApJ...763..130L}, which we find is due to flux
at $\approx$1.25\,$\mu$m dropping by 5$\times$ while flux at
$\approx$1.05\,$\mu$m only drops by 2.5$\times$. This behavior is
consistent with prior speculation that Y~dwarfs may be so cool that
the alkali atoms that dominate absorption at blue wavelengths for
warmer brown dwarfs finally become locked into molecules like Na$_2$S
and KCl, thereby reducing the opacity at 1.05\,$\mu$m relative to
1.25\,$\mu$m \cite{2012ApJ...758...57L}. The appearance of such
molecules could result in the return of substantial condensate clouds
\cite{2012ApJ...756..172M} and corresponding variability/weather.

In contrast to their near-infrared behavior, Y~dwarfs show remarkable
diversity in their mid-infrared colors. Even though they are only
$\approx$2$\times$ fainter than the latest T~dwarfs at these
wavelengths, they range from the same color as late-T dwarfs to much
redder ($\approx$0.8 magnitudes). One of the reddest objects is
WISEP~J1405+5534, which has been typed as ``Y0 peculiar?'' because its
$H$-band spectral peak is shifted 60\,\AA\ redder than the Y0 standard
WISEP~J1738+2732 \cite{2011ApJ...743...50C}. We find that
WISEP~J1405+5534 in fact has a very similar temperature to other
Y0~dwarfs (Table~S5) indicating that its unusual spectrum is due to
another physical property. Both the mid-infrared color and peculiar
spectrum may be explained by a reduced level of nonequilibrium
chemistry in the photosphere, perhaps due to reduced vertical
mixing. This would produce enhanced NH$_3$ absorption at $H$-band as
compared to other Y0~dwarfs and enhanced CH$_4$ absorption relative to
CO driving WISEP~J1405+5534 to redder $[3.6]-[4.5]$ colors.

\begin{figure}

\centerline{\includegraphics[width=4.75in,angle=0]{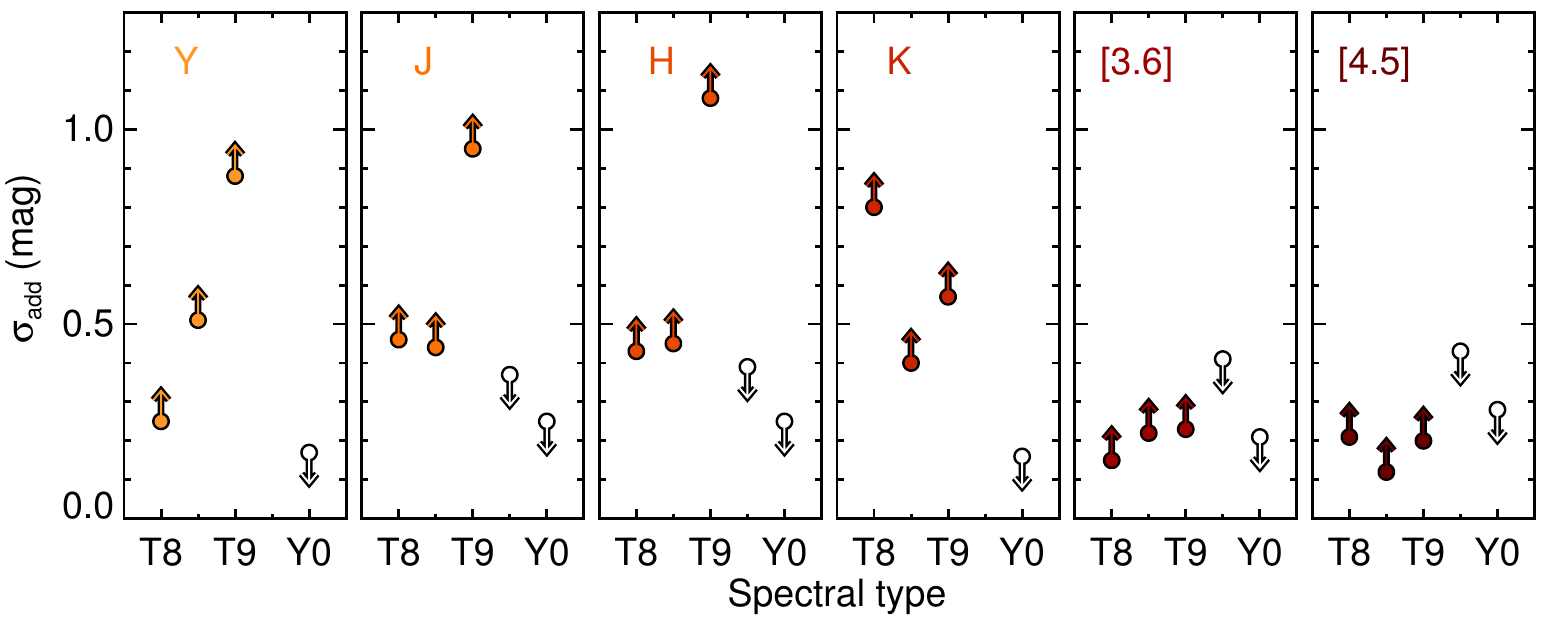}}

\caption{\normalsize Intrinsic photometric scatter among objects at
  each spectral type; unfilled symbols are upper limits where no
  scatter is detected. Y0~dwarfs show remarkably low intrinsic scatter
  in the near-infrared, reversing the trend observed at the end of the
  T~dwarf sequence that later type T~dwarfs show increasing dispersion
  in their near-infrared absolute magnitudes.  The much smaller sample
  of T9.5 dwarfs are also suggestive of this reversal at $J$ and $H$
  bands.  There is no evidence for such a reversal in the mid-infrared
  where both late-T and Y~dwarfs show much less intrinsic scatter and
  distance uncertainties for Y~dwarfs do not permit strong enough
  upper limits. \label{fig:rms}}

\end{figure}

The coldest brown dwarfs also demonstrate unusual behavior in their
absolute fluxes as a function of spectral type. Despite the plummeting
near-infrared flux---normal Y0~dwarfs are $\approx$6$\times$ fainter
than the latest T~dwarfs---Y0~dwarfs have indistinguishable fluxes
compared to each other to within 15\%--25\%. This is very unusual
compared to warmer brown dwarfs, which do not show such step-function
behavior at any spectral type transition and also show much larger
intrinsic scatter ($\approx$30\%--50\%) in absolute fluxes for a given
spectral type \cite{2012ApJS..201...19D}. This homogeneity among the
Y0~dwarfs is further unexpected because it reverses the trend observed
for the late-T dwarfs that the scatter increases substantially with
later type, cooler objects (Fig.~\ref{fig:rms}). For example, here we
double the sample of T9~dwarfs with accurate distances and find that
their near-infrared fluxes typically have a scatter of 130\%--210\%.

Another unexpected result is that T9.5 dwarfs appear to be brighter at
all bandpasses than the mean for T9 dwarfs and the T9 standard
UGPS~J0722$-$0540. Given the smaller sample of T9.5 dwarfs (three
objects) and their more uncertain distances, this brightening is
currently a 2$\sigma$ result, i.e., the weighted means in Table~S3 are
consistent with being equal at a $p$-value of 0.05.  Such a
brightening is reminiscent of the change in near-infrared fluxes from
late-L to early-T dwarfs
\cite{2006ApJS..166..585B,2006ApJ...647.1393L}, however we note that
the brightening at the L/T transition only occurs at blue
near-infrared wavelengths whereas we see brightening at all bands for
the T9.5 dwarfs.

To derive bolometric luminosities from the absolute fluxes, we computed
``super-magnitudes'' by summing the fluxes in near- and mid-infrared
bandpasses. This is an approximation to the standard method of
integrating the observed spectral energy distribution as a function of
wavelength, which is not possible for Y~dwarfs given the current lack
of sensitive mid-infrared spectrographs. We derived a multiplicative
correction to account for the remaining flux not captured in these
bands from a large grid of model atmospheres
\cite{2012ApJ...756..172M,2012ApJ...750...74S}.  The weak dependence
on these models is highlighted by the 8\% fractional uncertainty in
this correction factor (Fig.~S\ref{fig:bc}).

\begin{figure}

\centerline{\includegraphics[height=2.25in,angle=0]{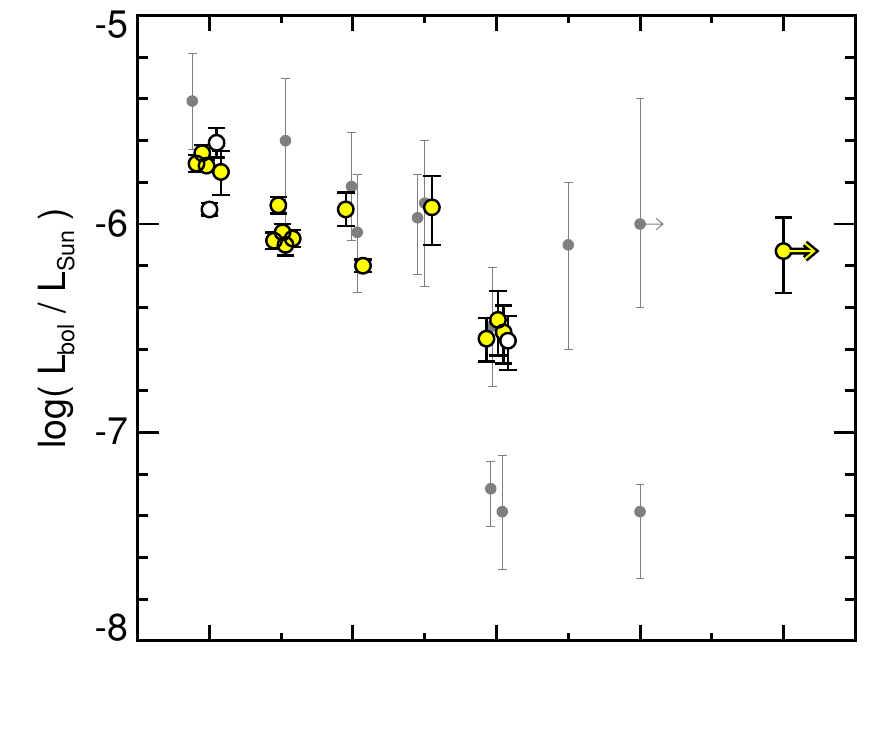}}
\vskip -0.30in
\centerline{\includegraphics[height=2.25in,angle=0]{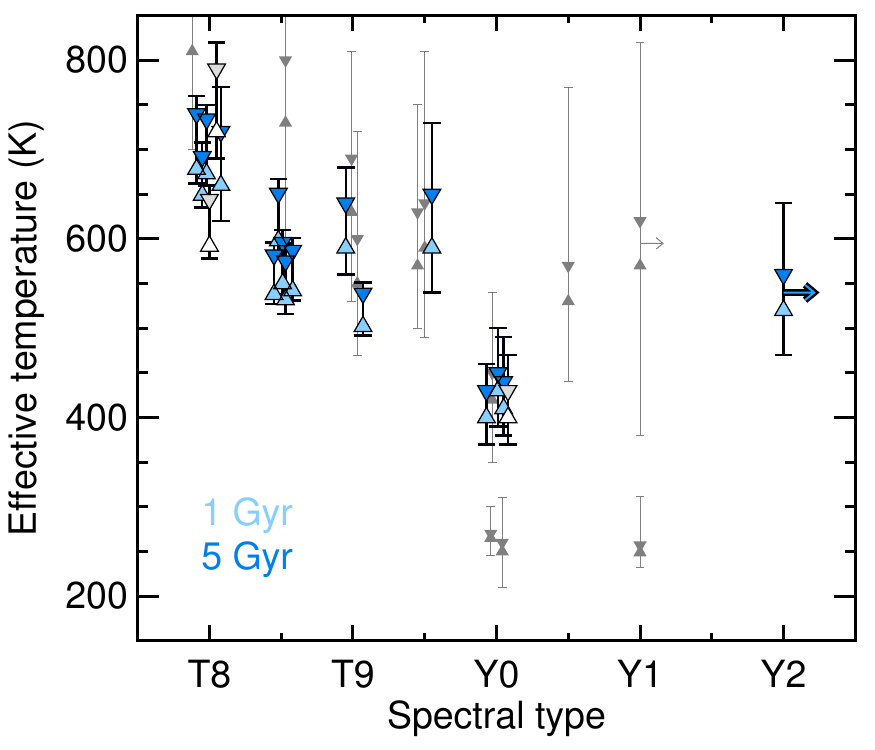}}

\caption{\normalsize Bolometric luminosities (\Lbol) and effective
  temperatures for objects of spectral type T8 and later; spectrally
  peculiar objects are denoted by white symbols.  Objects with \Lbol\
  uncertainties larger than 0.2\,dex are shown as smaller, gray
  symbols.  These are either objects with very uncertain distances or
  the components of tight binaries where the lack of resolved
  mid-infrared photometry results in a very uncertain bolometric flux.
  Error bars for spectral types are not plotted, and small $x$-axis
  offsets have been added to the spectral types for clarity.
  Effective temperatures are derived from our \Lbol\ measurements and
  Cond evolutionary model radii. Upward and downward pointing
  triangles correspond to the median \Lbol\ and lower and upper age
  limits used (see Table~S5). Error bars show the range of
  temperatures corresponding to the $\pm$1$\sigma$ range of \Lbol\
  over the same age range.  \label{fig:lbol-teff}}

\end{figure}

We used the Cond evolutionary models \cite{2003A&A...402..701B} to
estimate radii and thereby temperatures, masses, and surface gravities
from the bolometric luminosities of our sample
(Fig.~\ref{fig:lbol-teff}, Table~S5). We assumed fiducial ages of
1\,Gyr and 5\,Gyr as expected for the field population
\cite{2004ApJS..155..191B,2005ApJ...625..385A}. The tangential
velocities for our sample are consistent with having such typical
ages.  We find that the fractional change in temperature over this
narrow range of spectral types is remarkably large: the mean
temperature of T8~dwarfs is 685--745\,K (for 1--5\,Gyr), and this
drops to 410--440\,K for Y0~dwarfs (Table~S6).  Thus, these two
subtypes alone span the same fractional range in temperature as the
entire sequence of FGK stars (7300--4400\,K) that are $\sim$10$\times$
hotter.

Although much cooler than their late-T counterparts, Y0~dwarfs turn
out to be significantly warmer than previously suggested from model
atmosphere fitting (Fig.~S\ref{fig:teff-teff}). The most common
best-fit models of Y0~dwarfs in previous work have $\Teff = 350$\,K,
with plausible model fits of 400\,K in some cases
\cite{2011ApJ...743...50C}. Thus, model fits are typically 60--90\,K
($\approx$15\%--25\%) cooler than we find from our distances combined
with evolutionary model radii. If the fault lies with our assumed
radii, they would need to be 30\%--50\% larger than expected because
$\Teff \propto \Rstar^{-1/2}$. This would require very young ages
($\lesssim$100\,Myr) or very large systematic errors in the
evolutionary models that are not likely given the aforementioned
empirical validation from transiting brown dwarfs. Rather, we suggest
that parameters derived from fitting model atmospheres to
near-infrared spectra, where $\lesssim$5\% of the flux emerges, are
less likely to be accurate because current atmospheres imperfectly
reproduce observed spectra.

Using our luminosity measurements, we find that the coldest brown
dwarfs would be 6--10 Jupiter masses given an age of 1\,Gyr. An older
age of 5\,Gyr implies 16--25 Jupiter masses.  These masses therefore
straddle the current demarcation of ``planetary mass'' set by the
deuterium-fusing mass limit of $\approx$13 Jupiter masses
\cite{1996ApJ...460..993S,2000ARA&A..38..337C,2012A&A...547A.105M}.
Thus, it is possible that the atmospheres of our objects harbor
deuterated molecules such as HDO or CH$_3$D that have not yet been
detected because of the observational challenges
\cite{2000ApJ...542L.119C}.

Given the interest in both identifying the coldest atmospheric
benchmarks and searching for the bottom of the initial mass function
we briefly consider the most extreme objects in our sample in terms of
temperature and mass. WISEP~J1828+2650 has been dubbed the archetypal
Y~dwarf with a model-atmosphere temperature of $\lesssim$300\,K, i.e.,
room temperature, based on extremely red colors implying that the Wien
tail of its underlying blackbody distribution has moved into the
near-infrared \cite{2011ApJ...743...50C}. Our luminosity for this
object is inconsistent with such a low temperature, and we find it
must be at least 420\,K at 2$\sigma$; our calculations give
$520^{+60}_{-50}$\,K at an age of 1\,Gyr. (If WISEP~J1828+2650 is a
binary as proposed by \cite{2013ApJ...763..130L} the 2$\sigma$ limit
at 1\,Gyr only drops to 360\,K and 340\,K for the hypothetical two
components.) Its atypical properties compared to other Y~dwarfs may
simply be due to a slightly lower surface gravity, i.e., slightly
younger age, which qualitatively agrees with model predictions that
the collapse in near-infrared flux happens at warmer temperatures for
lower surface gravity \cite{2003ApJ...596..587B}. Ross~458C is a
contender for the lowest mass object, at 7 Jupiter masses, if its age
is near the 150\,Myr lower limit of its proposed age range
\cite{2010ApJ...725.1405B}. However, WD~0806$-$661B is the most secure
case for both lowest temperature (330--375\,K) and lowest mass (6--10
Jupiter masses) object known, given that it has a precise age of
$2.0\pm0.5$\,Gyr \cite{2012ApJ...744..135L}.

Overall, our results strengthen the connection between the coolest
brown dwarfs and gas-giant exoplanets. We validate that they probe an
extreme physical regime that bridges the gap between previously known,
hotter brown dwarfs and Jupiter-like planets.
We find that objects of very similar temperatures can have widely
varying spectral energy distributions and absorption features, e.g., a
range of 0.8~magnitudes in mid-IR colors for the same \Teff.  Along
with the fact that the $\geq$Y2~dwarf is warmer than the Y0~dwarfs,
this implies that temperature is not the principal determinant in
shaping spectra but rather seems to be on comparable footing with
other physical properties such as surface gravity, vertical mixing,
clouds, and perhaps metallicity. Consequently, the current spectral
classification scheme used to identify Y~dwarfs may not strongly
correlate with temperature as it generally does for L and
T~dwarfs. This could explain the unusually homogeneous fluxes for
Y0~dwarfs, unusually heterogeneous fluxes for T9~dwarfs, and plateau
or brightening of flux from T9 to T9.5.


\bibliographystyle{Science}


\begin{scilastnote}
\item We thank M.\ C.\ Liu for many helpful suggestions for improving
  this article; C.\ V.\ Morley for fruitful discussions regarding her
  and collaborators' model atmospheres, particularly the Vega zero
  points; M.\ C.\ Cushing for making available published spectra of
  our targets; and K.\ N.\ Allers for assistance in computing
  photometry from our IRAC data.
  This work is based on observations made with the \textsl{Spitzer
    Space Telescope}, which is operated by the Jet Propulsion
  Laboratory, California Institute of Technology under a contract with
  NASA.  
  T.J.D.\ acknowledges support from Hubble Fellowship grant
  HST-HF-51271.01-A awarded by the Space Telescope Science Institute,
  which is operated by AURA for NASA, under contract NAS 5-26555.
  A.L.K.\ was supported by a Clay Fellowship.
  Our imaging data are available in the \Spitzer\ Heritage Archive at
  \url{http://sha.ipac.caltech.edu/applications/Spitzer/SHA}, and the
  astrometric measurements we derived are given in Table~S1.

\end{scilastnote}



\clearpage

\clearpage

\clearpage

\clearpage

\clearpage

\includepdf[pages={2}]{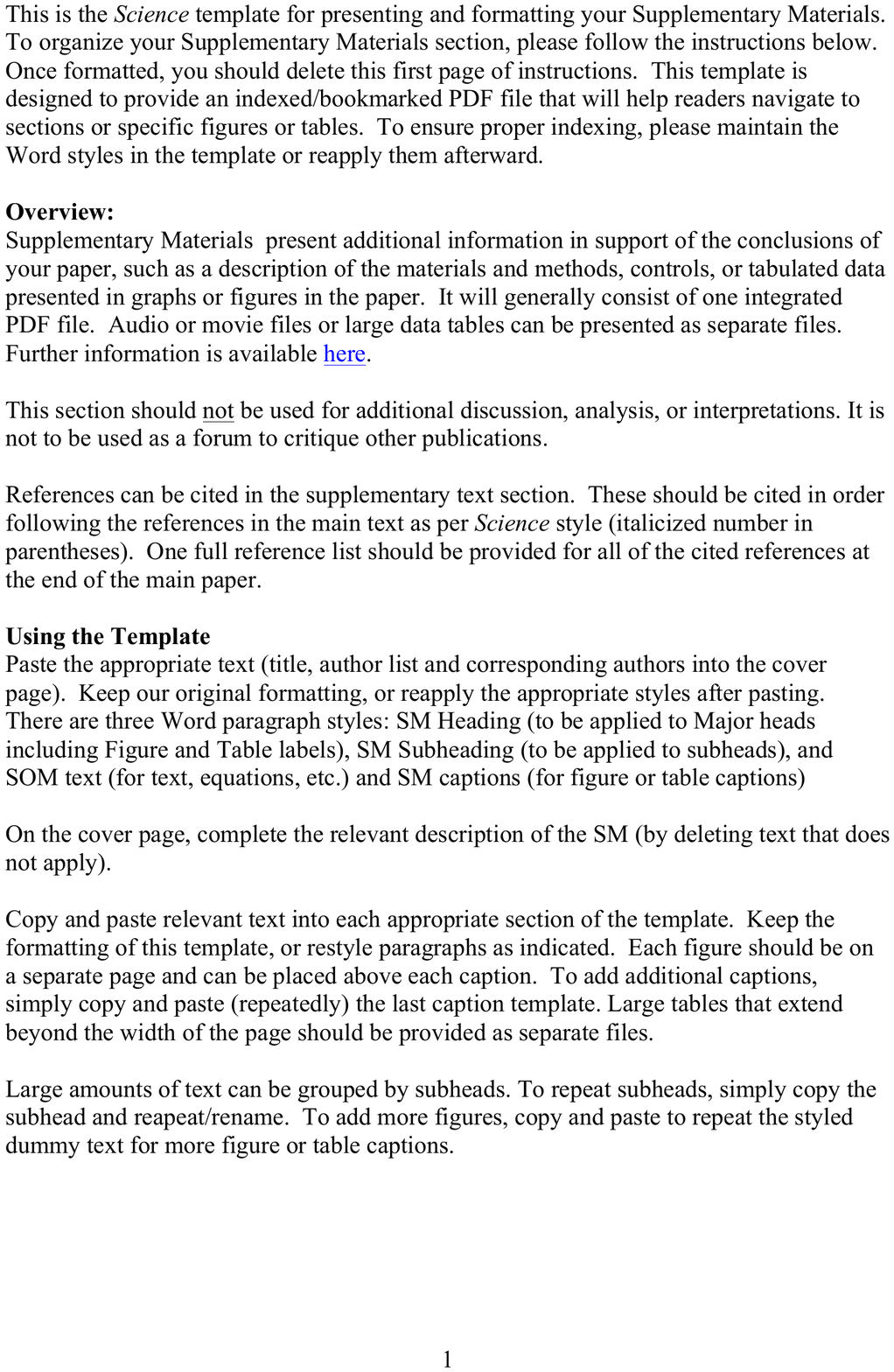}

\clearpage

\makeatletter
\makeatletter \renewcommand{\fnum@figure}
{\figurename~S\thefigure}
\makeatother
\setcounter{figure}{0}
\renewcommand{\figurename}{Figure}

\setcounter{page}{2}

\noindent \textbf{Materials and Methods}

\noindent \underline{Astrometric Monitoring with \Spitzer}

We used the Infrared Array Camera (IRAC) aboard the \textsl{Spitzer
  Space Telescope} \cite{2004ApJS..154...10F} in our astrometric
monitoring program targeting 16 late-T and Y~dwarfs. The data we
present here were obtained between 2011~November and 2012~December as
part of a Cycle~8 Director's Discretionary Time program (PID-80233).
Our target sample comprises eleven objects that did not have
previously published parallaxes, including all six Y~dwarfs known in
2011 \cite{2011ApJ...743...50C}, and five late-T dwarfs with parallax
measurements in the literature that serve as a check on our methods.

In the post-cryo (``warm'') \Spitzer\ mission, only IRAC channels~1
and 2 at 3.6\,$\mu$m and 4.5\,$\mu$m, respectively, are operational.
Each channel has its own $256 \times 256$ InSb detector with a pixel
scale of 1$\farcs$2\,pixel$^{-1}$, yielding fields of view of
$5\farcm2 \times 5\farcm2$. We chose to use channel~1, because of the
much larger number of reference stars available at shorter
wavelengths. The coldest Y~dwarfs have extremely red $[3.6]-[4.5]$
colors due to increased CH$_4$ opacity at 3.6\,$\mu$m and lowered CO
opacity at 4.5\,$\mu$m. However, even our faintest target at
3.6\,$\mu$m is still bright enough to be detected at high
signal-to-noise ratio (${\rm S/N} \gtrsim 20$) in a single 100-second
IRAC exposure.

We designed our \Spitzer\ monitoring program such that each target
would be observed at five epochs spanning $\gtrsim$1.5~years, as this
time baseline allows us to robustly disentangle parallax from proper
motion. 
All targets had previously been observed at least once with
\Spitzer/IRAC $\gtrsim$1~year before our program started, enabling us
to achieve the needed time baseline and number of epochs with four new
observations during Cycle~8. For most of our targets (13 of 16), only
two \Spitzer\ visibility windows per year are available, each lasting
$\approx$40--70~days. For several of these targets, three visibility
windows occurred during our program enabling us to obtain one epoch
each at the first and last window and two epochs during the middle
window. For the remainder we simply obtained two epochs per visibility
window. Three targets have one single $\approx$220-day window each
year, and for these we obtained epochs spaced uniformly in time.
For one target, the T8 dwarf PSO~J043.5395+02.3995, we lost a
2012~March epoch due to the extremely large number of radiation hits
caused by a solar storm, but the other epoch during that visibility
window was unaffected.

\noindent \underline{\Spitzer\ Astrometry Pipeline}

At each epoch we obtained 9, 18, or 36~dithered images, with more
images for fainter targets in order to increase the final S/N of their
mean positional measurements.  For all of the following analysis we
use the ``corrected basic calibrated data'' products from the
automated \Spitzer\ pipeline processing.  These low-level products
have standard corrections applied for detector bias, nonlinearity,
pixel-to-pixel response (i.e., flat fielding), and well-understood
image artifacts such as column pulldown and muxstripes.

We obtained positional measurements from all sources in each field
from the implementation of DAOPHOT \cite{1987PASP...99..191S} in IRAF.
The point-spread function (PSF) model used by DAOPHOT was defined
within a radius of 8$\arcsec$, and we used a fitting radius of
5$\arcsec$, i.e., 3.0\,$\times$ the 1$\farcs$7 full-width at
half-maximum (FWHM) of IRAC channel~1 images. For \texttt{daofind}, we
allowed very liberal sharpness and roundness thresholds since the
undersampled IRAC images do not constrain these parameters well ($0 <
{\rm \tt sharp} < 99$; $-9 < {\rm \tt round} < 9$), and spurious
detections were rejected by clipping later in our analysis. We then
used \texttt{phot} to measure positions and fluxes for each of the
sources from \texttt{daofind}, using an aperture of 2$\farcs$5 and a
sky annulus of 3$\farcs$3--5$\farcs$8. Positions were generated by the
default centroiding algorithm using a centering box of 5$\arcsec$. We
applied our IRAC distortion solution to the resulting $(x,y)$
positions directly within IRAF using the routine \texttt{xygeotran},
since this is the native environment in which we measured and stored
the polynomial coefficients.

We analyzed the resulting position measurements in nearly an identical
fashion as in our previous work using ground-based infrared imaging
from the Canada-France-Hawaii Telescope \cite{2012ApJS..201...19D}.
First, we created an astrometric catalog at each epoch by
cross-identifying detections and registering the individual
dithers. At this initial stage we excluded the lowest S/N detections,
applying a threshold in S/N that ranged from 5--10 depending on the
data set. (For denser fields we could afford stricter cuts.) Our
method for cross-identifying sources used a temporary astrometric
solution for the field, created using the \WISE\ All-Sky Source
Catalog \cite{2010AJ....140.1868W}. We only kept sources that are
detected in $\geq$50\% of our frames, and we $\sigma$-clipped these
measurements, both of which effectively eliminate spurious detections
from appearing in the final astrometric catalog for a given epoch. For
positional uncertainties, we used the standard error of the
measurements. Next, we registered the astrometry between epochs,
masking all sources with large proper motion ($>$100\,\masyr),
including the target, during this process. Finally, we determined the
absolute astrometric calibration (e.g., pixel scale and orientation)
by matching sources with low proper motion to the \WISE\ All-Sky
Source Catalog.

The properties of each astrometric catalog are given in Table~S1,
which lists the number of epochs, time baseline, and total number of
reference stars as well as the number of reference stars matched with
the \WISE\ catalog. For our observations, the median astrometric
precision per epoch for our targets was 30\,mas, with 97\% between
20\,mas and 40\,mas. Our median target S/N was 27, while the first
epoch archival data was sometimes of lower S/N and thus had somewhat
larger astrometric errors (median of 50\,mas and 90\% were
$<$60\,mas).

\noindent \underline{Parallaxes and Proper Motions}

We determined the proper motions and parallaxes of our targets using
essentially the same method as described in Section~2.4 of
\cite{2012ApJS..201...19D}. We found the best-fit solution using MPFIT
in IDL \cite{2009ASPC..411..251M} and then performed a Markov Chain
Monte Carlo (MCMC) analysis using 30 chains each with 10$^6$ steps.
The key difference is that we used the JPL ephemeris for \Spitzer\
rather than the JPL DE405 ephemeris of the Earth when computing
parallax ellipses. 
In Table~S1 we give the parameters derived from our MCMC
analysis: right ascension ($\alpha$), declination ($\delta$), parallax
($\pi_{\rm rel}$), and proper motion ($\mu_{\alpha}\cos{\delta}$ and
$\mu_{\delta}$). The resulting parameter distributions all appear to
be consistent with Gaussians based on fits to their histograms, so we
simply quote the median and standard deviation derived from each set
of chains. Note that the parallax and proper motion are relative in
the sense that the astrometric reference frame for each target is
defined by stars at a finite distance and thus has some mean parallax
and proper motion. However, even in our shallowest 12-second
exposures, the Besan\c{c}on model of the Galaxy
\cite{2003A&A...409..523R} predicts that the mean parallax of our
reference stars is 1.5\,mas and that 90\% of the stars have parallaxes
$<$3\,mas. This is $\approx$7--20$\times$ smaller than our parallax
errors and thus negligible. The $\chi^2$ of our best-fit solutions are
all commensurate with the degrees of freedom (${\rm dof} = 2 \times
N_{\rm epoch} - 5$). This validates our assumed positional
uncertainties and derived parameter errors. The best-fit parallax
solutions are shown in Fig.~S\ref{fig:plx}.

The median parallax precision for our entire sample is 17\,mas, and
the median fractional uncertainty is 16\% or S/N = 6.2. For 11 of the
16 targets the parallax uncertainty is $\leq$20\% or ${\rm S/N} \geq
5.0$. These parallax errors are currently limited by the available
time baseline, not the per epoch astrometric precision, i.e., if the
same number and quality of measurements were spread out over more time
then the errors would decrease. One control target 2MASS~J0415$-$0935
(T8) has a very long time baseline (8.27~years), as its earliest
observations date from the first few months of the \Spitzer\ mission.
It has similar per epoch precision as other targets but the smallest
parallax uncertainty (10\,mas), illustrating the point that a long
time baseline enables a better determination of the proper motion and
thus parallax.

Finally, we note that we visually inspected images from the first and
last epochs to determine if our targets may have been blended with
background stars during any of our observations. The control T8p dwarf
2MASS~J0729$-$3954 appears in the first epoch to have been passing
very close to a star that is $\approx$1.3\,mag fainter at $[3.6]$.
Given this background star's position of $(\alpha, \delta) =
(112\fdg2488, -39\fdg8959)$, 2MASS~J0729$-$3954 would have passed as
close as 2$\farcs$7 to this star both in our data and in the
Blanco/ISPI imaging used by previous authors to measure its parallax
\cite{2012ApJ...752...56F}. Thus, both of our parallax measurements
may be biased by contaminating light from this background star.

\noindent \underline{Comparison to Published Parallaxes \& Lutz-Kelker Bias}

Our five control targets have parallaxes that are in good agreement
with published values, providing validation of our methods. The
$\chi^2$ of differences between our parallax values and those
previously published is 5.7 (5 dof, $p=0.34$). The largest discrepancy
is for 2MASS~J0729$-$3954 for which previous work found $\pi =
126\pm8$\,mas \cite{2012ApJ...752...56F} and we find $91\pm25$\,mas,
which is only a 1.4-$\sigma$ difference and may be due to the
contaminating light from a nearby star as noted above. Our other
results typically agree within 1$\sigma$ of published values
\cite{2012ApJS..201...19D,2012ApJ...748...74L,2011ApJ...740L..32L,2007A&A...474..653V}. Our
proper motions typically also agree well, even though these are
relative measurements made in a different reference frame than
published values.
Most proper motions agree within 1$\sigma$, and the most discrepant
value $\mu_{\alpha}\cos{\delta}$ for 2MASSI~J0415$-$0935 (3.0$\sigma$;
10\,\masyr).
Seven of our science targets have recently published parallaxes from
\cite{2013ApJ...762..119M}, and the $\chi^2$ of differences between
their values and ours is 5.7 (7 dof, $p=0.58$). Thus, our parallaxes
are in good agreement with the results of \cite{2013ApJ...762..119M},
but our uncertainties are $\approx$2--4$\times$ smaller. Our smaller
errors are likely due to the fact that we are using a higher quality
distortion solution for \Spitzer/IRAC and $\sim$10$\times$ more
reference stars. We also note that most of our proper motion values
agree within 1$\sigma$ compared to \cite{2013ApJ...762..119M}, with
three being different by 1.1--2.0$\sigma$. A comparison of all our
parallaxes to published results is shown in Fig.~S\ref{fig:vs-pub}.

One of our science targets, WISEP~J1828+2650, has a few different
reported parallaxes in the literature. One value of $122\pm13$\,mas
\cite{2012ApJ...753..156K} was based on preliminary results later
published in \cite{2013ApJ...764..101B}, and this is 2.7-$\sigma$
discrepant with our value of $70\pm14$\,mas. These authors more
recently reported three values based on two different methods of
computing relative astrometry (see their Table~7). Their ``method 1''
gives $103\pm16$\,mas, 1.6-$\sigma$ larger than our value, and their
``method 2'' gives $79\pm12$\,mas in good agreement with our parallax
\cite{2013ApJ...764..101B}. A parallax of $90\pm9.5$\,mas is also
reported based on combining the two methods
\cite{2013ApJ...764..101B}, but we choose not to adopt this value as
it likely underestimates the measurement uncertainty. The two methods
use the same underlying imaging data and thus are not truly
independent data sets, so they cannot simply be combined to reduce the
measurement errors. In the following, we use our parallax value for
WISEP~J1828+2650 since it has a similar uncertainty to the
\cite{2013ApJ...764..101B} values but is based on data from a single
bandpass and telescope, which reduces the chances for systematic
errors. In particular, if WISEP~J1828+2650 is an unresolved binary
with components having different colors, as suggested by
\cite{2013ApJ...763..130L}, then its photocenter would shift between
different bandpasses.

Finally, we consider the handful of our targets with low significance
parallax detections. One science target WISEP~J1541$-$2250 (${\rm S/N}
= 2.4$; Y0.5) and one control object Ross~458C (${\rm S/N} = 2.4$; T8)
have parallax ${\rm S/N} < 3$. As we discuss above, our results for
the control objects are in good agreement with the more precise
published values. For WISEP~J1541$-$2250, our MCMC analysis gives an
upper limit of 148\,mas for the parallax at 99\% confidence. While our
results for this object agree with the new value of $-21\pm94$\,mas
\cite{2013ApJ...762..119M}, it is highly discrepant with the earlier
measurement from the same group of $351\pm108$\,mas
\cite{2011ApJS..197...19K}.

Two additional science targets, WISEP~J0148$-$7202 (${\rm S/N} = 3.8$;
T9.5) and WISEP~J0458+6434 (${\rm S/N} = 3.7$; T8.5+T9.5), and one
control target 2MASS~J0729$-$3954 (${\rm S/N} = 3.6$; T8p) have $3 <
{\rm S/N} < 5$ parallaxes. For cases of such low S/N, an assumption
that objects are distributed uniformly in space volume would naturally
lead to a strong prior in parallax and thereby result in a systematic
offset in measured values, i.e., ``Lutz-Kelker bias''
\cite{1973PASP...85..573L}. If other prior information is available,
e.g., about absolute magnitudes or velocities, this can be used to
mitigate Lutz-Kelker bias, particularly in the lowest S/N cases where
the very steep prior causes zero parallax solutions to dominate for
${\rm S/N} \lesssim 4$, the ``Lutz-Kelker catastrophe''
\cite{2003AJ....126.3017T}. However, we do not yet know the expected
brightness of Y~dwarfs or if their velocity distribution is different
from better studied L and T~dwarfs.

Instead, we have investigated the validity of the uniform volume prior
given that our science targets were discovered in an all-sky
magnitude-limited survey. We simulated this assuming that the
underlying absolute magnitudes followed a normal distribution and that
objects were distributed uniformly in space.  The resulting distance
distributions when applying a selection cut in apparent magnitude are
shown in Fig.~S\ref{fig:lutz}.  These are all essentially lognormal
distributions of differing widths that have a tail at small distances
that matches a uniform volume prior.  Thus, the slope of the prior,
i.e., whether large or small distances are preferred, actually depends
on where the object is with respect to the magnitude limit of the
survey, and this would require a prior assumption for its absolute
magnitude.  If most objects are found near the survey limits, as is
usually the case, a roughly flat prior corresponding to the peaks of
these distributions would actually be most appropriate.  Since we do
not know the absolute magnitudes Y~dwarfs a priori, we conservatively
choose to adopt a simple, uniform prior in the parallax. This approach
is supported by a test using the two control objects with low
significance parallaxes. For 2MASS~J0729$-$3954 and Ross~458C we tried
a uniform volume prior in our MCMC analysis by adding
$4\log(\pi/\pi_{\rm best\textnormal{-}fit})$ to the $\chi^2$ and found that the
resulting parallaxes were brought out of agreement with literature
values under such a prior. 


\noindent \underline{Absolute Magnitudes \label{sec:absmag}}

In Table~S2, we have compiled the available near- and mid-infrared
photometry for our sample as well as all other objects of spectral
types T8 or later with distance measurements. For the mid-infrared we
use \Spitzer/IRAC photometry since it is typically of much higher S/N
than \WISE\ catalog photometry for the latest type sources, which are
often not detected in the $W1$ band that is similar to IRAC's $[3.6]$
band. We have supplemented \Spitzer\ photometry from the literature by
performing aperture photometry on archive images of PSO~J043.5+02.39
(T8). The only other objects without \Spitzer\ photometry are
WISE~J1639$-$6847 (Y0:) and the components of the tight binaries
2MASSW~J1225$-$2739 (T5.5+T8), WISEP~J0458+6434 (T8.5+T9.5),
WISEP~J1217+1626 (T9+Y0), and CFBDSIR~J1458+1013 (T9+Y?) that are
unresolved in \Spitzer\ images. For near-infrared photometry we use
the results of \cite{2013ApJ...763..130L} where possible for the
Y~dwarfs, otherwise photometry from
\cite{2011ApJS..197...19K,2012ApJ...753..156K}. These published
results are all on the Mauna Kea Observatories (MKO) photometric
system, except for WISEP~J1741+2553, which is on the 2MASS system. We
computed MKO--2MASS offsets of $J_{\rm MKO} - J_{\rm 2MASS} =
-0.30$\,mag, $H_{\rm MKO} - H_{\rm 2MASS} = 0.07$\,mag, and $K_{\rm
  MKO} - K_{s,{\rm 2MASS}} = 0.13$\,mag for this object from its
Magellan/FIRE spectrum published by \cite{2011ApJS..197...19K}. The
resulting color--magnitude diagrams are shown in Fig.~\ref{fig:cmd}
where we include earlier type field dwarfs from our Database of
Ultracool Parallaxes for context.\footnote{Maintained by T.\ Dupuy at
  \url{http://www.cfa.harvard.edu/~tdupuy/plx}; updated 2012-06-09.}

In Table~S3, we give mean absolute magnitudes for each spectral type
bin, along with a lower or upper limit on the amount of intrinsic
scatter in each bin, depending on whether the rms in the magnitudes is
consistent with that expected from measurement uncertainties or not,
i.e., $p(\chi^2) \geq 0.5$ or $p(\chi^2) < 0.5$, respectively. Note
that we chose a $p$-value cutoffs of 0.5/0.5 here rather than
0.95/0.05 or 0.05/0.95 since we do not know a priori whether to expect
significant intrinsic scatter or not. The null hypothesis is not
necessarily that there should be zero intrinsic scatter, since earlier
spectral types often (but not always) show significant scatter in
absolute magnitudes \cite{2012ApJS..201...19D}. These upper/lower
limits on the intrinsic photometric scatter at each spectral type are
shown in Fig.~\ref{fig:rms}.
We excluded from our table of mean magnitudes and the discussion here
two Y0 dwarfs that have unreliable parallaxes, WISE~J0359$-$5401 and
WISE~J1639$-$6847.

In our analysis we use the tabulated mean absolute magnitudes rather
than polynomial fits as a function of spectral type that are commonly
used in other work \cite{2012ApJ...753..156K,2013ApJ...762..119M}.
This is because smooth polynomials often do not accurately capture
changes in absolute flux with spectral type. At spectral types of T8,
T8.5, and T9, our mean absolute magnitudes are within $\pm$0.3\,mag of
the polynomial relation between $H$-band absolute magnitude and
spectral type that excludes WISEP~J1828+2650 ($\geq$Y2) from
\cite{2012ApJ...753..156K}. Since we find that T9.5 dwarfs are
brighter than Y0 dwarfs, which impossible to capture with the smooth
polynomial from \cite{2012ApJ...753..156K}, our mean value is 1.8\,mag
brighter than their relation at this spectral type. Finally, our mean
$H$-band fluxes for Y0~dwarfs are 0.6\,mag brighter than the
polynomial from \cite{2012ApJ...753..156K}. This is simply due to the
polynomial undershooting their data points that happen to be very
similar to ours, despite updated distances since their preliminary
parallaxes and improved photometry from \cite{2013ApJ...763..130L}.


\noindent \underline{Calculating Bolometric Luminosities}

As little as $\lesssim$3\% of the emergent flux of Y~dwarfs is
expected to be emitted in the standard near-infrared windows at $YJHK$, and
no existing facilities are capable of measuring the spectra of
Y~dwarfs in the mid-infrared where they emit most of their light. Therefore
we must rely on models to some extent when deriving bolometric
luminosities for our sample, since we cannot directly integrate the
observed SEDs. Fortunately, the available photometry typically
captures $\gtrsim$50\% of the bolometric flux, which helps weaken this
dependence on models.

Rather than use a single bandpass, we develop a method of summing the
fluxes from individual bandpasses to compute ``super-magnitudes'' that
combine both mid-infrared and near-infrared flux when possible. To
have as much uniformity in our \Lbol\ calculations as possible, we
chose filter combinations for which the largest subsets of targets
have available photometry. All single objects earlier than Y1 in our
sample have photometry in $J$, $H$, $[3.6]$, and $[4.5]$ bands, so
this defined the main super-magnitude we used
($m_{JH12}$).\footnote{The Vega zero points we used to convert
  magnitudes into fluxes for $\{Y,J,H,K,[3.6],[4.5]\}$ bands were,
  respectively, $\{5.690, 4.322, 3.139, 1.318, 0.456, 0.235\} \times
  10^{-7}$\,erg\,cm$^{-2}$\,s$^{-1}$. The zero-points for our
  super-magnitudes are simply the sum of the Vega fluxes, e.g., for
  $m_{JH12}$ this would be $8.152 \times
  10^{-7}$\,erg\,cm$^{-2}$\,s$^{-1}$.} Note that adding $K$-band data
would not contribute much additional flux, since it typically contains
$\lesssim$10\% the total $J$+$H$ flux for such cool objects. At the
very latest types ($\geq$Y1), near-infrared photometry is usually not
available so we used only a sum of the $[3.6]$- and $[4.5]$-band flux
($m_{12}$).

We used model atmospheres
\cite{2012ApJ...756..172M,2012ApJ...750...74S} to compute bolometric
corrections for the super-magnitudes $m_{JH12}$ and $m_{12}$. To span
the possible range of properties for our sample, we used models with
temperatures of \{$300, 400, 500, 600, 700$\}\,K and surface gravities
of $\{1, 3, 10\} \times 10^4$\,cm\,s$^{-2}$. We used models with cloud
sedimentation parameters ranging from $f_{\rm sed} = 2$--5, i.e.,
thick to thin clouds \cite{2012ApJ...756..172M}, as well as
corresponding cloud-free models \cite{2012ApJ...750...74S}. Fig.~9 of
\cite{2012ApJ...756..172M} shows that at gravities of 10$^4$ and $3
\times 10^4$\,cm\,s$^{-2}$ the $f_{\rm sed} = 3$, 4, 5, and cloud-free
models agree best with the properties of late-T dwarfs on
near-infrared color magnitudes, and for $g = 10^5$\,cm\,s$^{-2}$ the
$f_{\rm sed} = 4$, 5, and cloud-free models agree best. Therefore, we
used only these subsets of cloud parameters, and this resulted in 47
different models being used to calculate bolometric corrections. We
took the mean and standard deviation of the derived values and found
${\rm BC}_{JH12} = 2.93\pm0.08$\,mag and ${\rm BC}_{12} =
5.29\pm0.21$\,mag (Fig.~S\ref{fig:bc}). (Note that we initially tried
larger ranges of model parameters but found that they did not
significantly change the resulting bolometric corrections, since most
of the flux is already captured by the super-magnitudes.)  For future
reference, we also computed bolometric corrections using bands 1 and 2
of \WISE\ instead of IRAC.  We calculate a $J$+$H$+$W1$+$W2$
bolometric correction of $2.93\pm0.06$\,mag and $W1$+$W2$ bolometric
correction of $5.12\pm0.21$\,mag, assuming model magnitude zero points
of $5.751 \times 10{-8}$\,erg\,cm$^{-2}$\,s$^{-1}$ for $W1$ and $2.527
\times 10{-8}$\,erg\,cm$^{-2}$\,s$^{-1}$ for $W2$.

In order to derive bolometric luminosities for objects without
mid-infrared photometry, mostly components of tight binaries, we
derived bolometric corrections for near-infrared magnitudes alone.
Rather than rely directly on models, we used the apparent bolometric
magnitudes ($m_{\rm bol}$) of the single objects that have
mid-infrared photometry to compute bolometric corrections at $Y$, $J$,
$H$, and super-magnitudes of $Y$+$J$, $J$+$H$, and $Y$+$J$+$H$. We
then took the weighted average at each spectral type, and these values
are reported in Table~S4. The rms about these mean values was 0.6\,mag
for all filter combinations, and we consider this to be the
uncertainty in these bolometric corrections. As expected, the scatter
is much higher without mid-infrared photometry, since even $Y$+$J$+$H$
captures only $\approx$10\% of the bolometric flux. However, this
method arrives at much more precise bolometric corrections than are
possible using models alone (e.g., if we used models as above but for
a $YJH$ super-magnitude we would derive ${\rm BC}_{YJH} =
0.9\pm1.2$\,mag).

In Table~S5 we list the apparent bolometric magnitudes derived for our
sample along with the final bolometric luminosities, where
$\log(\Lbol/\Lsun) \equiv (4.7554 - m_{\rm bol} + 5\log{d} - 5)/2.5$,
where $d$ is the distance in parsecs.\footnote{The bolometric absolute
  magnitude of the Sun is from
  \url{http://www.pas.rochester.edu/~emamajek/sun.txt}. Note that we
  also recompute the model $\log(\Lbol/\Lsun)$ values from radius and
  \Teff\ using the corresponding solar luminosity of $3.827 \times
  10^{33}$\,erg\,s$^{-1}$ for consistency.} We quote $m_{\rm bol}$ and
\Lbol\ separately so that improved parallax measurements in the future
can be readily applied to compute new luminosities. In
Fig.~\ref{fig:lbol-teff} we plot our derived \Lbol\ values as a
function of spectral type. As a test of our methods, we check the two
T8~dwarfs with published values for \Lbol\ based on near-infrared and
mid-infrared spectra. The published values of
$\log(\Lbol/\Lsun)=-5.67\pm0.02$\,dex for 2MASSI~J0415$-$0935
\cite{2007ApJ...656.1136S} and $-5.69\pm0.03$\,dex for
2MASS~J0939$-$2448 \cite{2008ApJ...689L..53B} are more precise than
our values and agree well with our derived luminosities, within
1.3$\sigma$ and 0.4$\sigma$, respectively. In the following analysis,
we use our \Lbol\ values for these two objects for consistency when
comparing results among the rest of the sample.


\noindent \underline{Deriving Effective Temperatures and Other Fundamental Properties}

Effective temperature is defined as $\Teff \equiv (4 \pi \sigma
\Rstar^2 / \Lbol)^{-1/4}$, where $\sigma$ is the Stefan--Boltzmann
constant. We can therefore derive effective temperatures for our
sample late-T and Y~dwarfs using our measured luminosities and an
assumption for their radii. 
Evolutionary models \cite{2000ApJ...542..464C,2003A&A...402..701B}
generally agree well with the measured radii of transiting substellar
objects over a wide range of masses from $\approx$5--60\,\Mjup\
\cite{2008A&A...491..889D,2009Natur.460.1098H,2011ApJ...726L..19A,2011A&A...528A..97T}.
We note that transiting objects in the $\approx$5--20\,\Mjup\ mass
range that we are most interested in may have formed via core
accretion or have been subjected to intense stellar irradiation over
their lifetimes, either of which could alter their radii compared to
the solar-abundance, gas-only evolutionary models relevant for our
sample.  However, despite the fact that they may not be ideal test
cases, the ensemble of measurements to date display the expected
trends that more intensely irradiated objects are inflated compared to
non-irradiated models \cite{2012ApJ...761..123S} and massive objects
most likely to be entirely gaseous and least likely to be affected by
irradiation agree well with models \cite{2011ApJ...730...79J}.
Furthermore, we note that observed variations in radii compared to
models are relatively small and similar to the variations predicted
over the range of plausible ages and masses.  Evolutionary models
fortuitously predict that the mass--radius relationship is nearly flat
with maximal variations of only $\pm$15\% over more than an order of
magnitude in mass (5--80\,\Mjup). At ages typical for field brown
dwarfs, radii are predicted to contract by only $\approx$5\% at a
given mass from 1\,Gyr to 5\,Gyr, although we note that for much
younger ages their radii can be substantially larger (e.g., 10\%--20\%
larger at 0.1\,Gyr relative to 1\,Gyr for masses of 5--20\,\Mjup).
Ultimately, any variations in radius have a comparatively small impact
on our derived temperatures since $\Teff \propto \Rstar^{-1/2}$.

We use the Cond evolutionary model isochrones
\cite{2003A&A...402..701B} at ages of 1\,Gyr and 5\,Gyr to derive
radii, temperatures, masses, surface gravities, and deuterium
abundances for our sample. We interpolate the logarithm of these
quantities from each isochrone as a function of $\log(\Lbol)$. We have
chosen these models because they are among the most widely used models
that are appropriate for objects that have no silicate condensate
clouds in the photosphere. In principle, clouds at earlier stages of
evolution can have some impact on properties at older ages, but the
available models accounting for such effects do not currently extend
to low enough luminosities \cite{2008ApJ...689.1327S}. New
evolutionary models are currently being developed that not only
account for silicate cloud evolution but previously neglected sulfide
clouds \cite{2012ApJ...756..172M}. As an example of the differences
between evolutionary models with varying boundary conditions and
interior structure physics, the predicted radii from
\cite{2008ApJ...689.1327S} are 3\%--5\% higher over the 5--30\,\Mjup\
mass range at 1\,Gyr compared to Cond models. This contributes a
negligible uncertainty of 1.5\%--2.5\% to our derived temperatures.

In Table~S5 we list the model-derived properties for each object, with
uncertainties given solely by the individual luminosity errors
propagated through the interpolation of the models. Note that these
error bars therefore only reflect the rms in the parallaxes and
bolometric corrections at a given age and do not include any potential
systematic errors in our bolometric corrections or in the evolutionary
model isochrones.
In Table~S6, we report weighted averages as a function
of spectral type for the luminosities and model-derived properties of
``normal'' objects. We do not report individual values of the
deuterium abundance relative to the initial abundance (D/D$_0$), since
it is typically either zero or unity. The Cond models predict that at
1\,Gyr objects should have retained nearly all of their initial
deuterium for $\Teff \leq 500$\,K and should have depleted almost all
of it for $\Teff \geq 615$\,K. At 5\,Gyr, Cond models predict that
deuterium boundary lies at $\Teff = 320$--390\,K, i.e., cooler than we
find for normal T8--Y0 dwarfs at that age. We note that this is a
potential test of the ages/masses of this sample, since an older,
i.e., higher mass, population of Y0~dwarfs should show no evidence of
deuterium, whereas younger objects of similar temperature would retain
most or all of their deuterium. 

A few objects in our sample are companions to more massive stars with
independent age constraints. Ross~458AB has an age in the range
150--800\,Myr based on strong chromospheric activity and a lack of
spectroscopic signatures of very low surface gravity
\cite{2010ApJ...725.1405B}. Thus, we used these ages instead of 1\,Gyr
and 5\,Gyr to derive properties from the Cond models. (Note that on
the extreme ends of this age range, Ross~458C is expected to retain
all or none of its initial deuterium.) WD~0806$-$661A has a white
dwarf cooling age of $2.0\pm0.5$\,Gyr \cite{2012ApJ...744..135L}, and
thus we use ages of 1.5\,Gyr and 2.5\,Gyr for WD~0806$-$661B. The
other companions, Wolf~940B at 3.5--6.0\,Gyr
\cite{2009MNRAS.395.1237B}, BD+01~2920B at 2.3--14.4\,Gyr
\cite{2012MNRAS.422.1922P}, and WISE~J1118+3125 at 2--8\,Gyr
\cite{2013AJ....145...84W}, have ages that are not well constrained
but are broadly consistent with one or both of our fiducial ages of
1\,Gyr and 5\,Gyr, so we do not calculate separate properties for
these objects.

In Fig.~\ref{fig:lbol-teff} we show our derived effective
temperatures as a function of spectral type. The mean luminosity for
T8~dwarfs is in good agreement with previous estimates based on more
extensive SED coverage, $\log(\Lbol/\Lsun) = -5.70$\,dex, and at ages
of 1\,Gyr and 5\,Gyr this corresponds to 685\,K and 745\,K,
respectively. Normal Y0~dwarfs have a mean temperature of 410\,K and
440\,K for assumed ages of 1\,Gyr and 5\,Gyr. They are more than an
order of magnitude less luminous than T8~dwarfs, and correspondingly
\Teff\ drops by 40\% and masses are predicted to be lower by a factor
of $\approx$2. (Note that evolutionary models predict that radii
increase by $\approx$20\% with decreasing mass over the range
$\approx$5--60\,\Mjup, so this slightly counteracts the trend of lower
luminosity objects at a given age having lower \Teff.)

The temperatures of Y0~dwarfs, despite being much cooler than their
late-T counterparts, are significantly warmer than found by model
atmosphere fitting \cite{2011ApJ...743...50C}. In
Fig.~S\ref{fig:teff-teff} we show these published temperatures for
objects in common with our parallax sample. For four ``normal''
Y0~dwarfs (including WISEP~J1541$-$2250, which is now classified as
Y0.5), model atmosphere best fits give 350\,K
\cite{2011ApJ...743...50C}. The \Teff\ range of plausible model fits
were 350--400\,K in two cases (350\,K only in the others) with
best-fit gravities ranging from $\logg = 3.75$ to 4.75 (cgs)
\cite{2011ApJ...743...50C}.

\bigskip
\bigskip

\noindent \textbf{Supplementary Text}

\noindent \underline{Tangential Velocities}

Proper motions combined with distance measurements directly yield
velocities in the tangent plane of the sky (\Vtan). For earlier type
late-M, L, and T~dwarfs, the median and rms is $\Vtan \approx
30\pm20$\,\kms\ \cite{2009AJ....137....1F} computed from samples of
$\sim$100 objects per spectral class with measured proper motions and
spectroscopic distance estimates. Two Y~dwarfs have published values
of $\Vtan \gtrsim 100$\,\kms\ \cite{2011ApJS..197...19K}, suggesting
that this population may have significantly higher \Vtan\ than
T~dwarfs.

For our entire sample of T8 or later objects, we find a weighted mean
$\Vtan = 23$\,\kms\ with an rms of 27\,\kms, which is generally
consistent with earlier type objects \cite{2009AJ....137....1F}. For
just the Y0~dwarfs, we find a weighted mean of $\Vtan = 45$\,\kms\ and
rms of 30\,\kms, both slightly higher than at earlier types but not as
high as initial published estimates \cite{2011ApJS..197...19K}. This
is partly because our parallax distances are 10\%--20\% closer than
earlier photometric distance estimates, but also because the earlier
proper motion precision was not sufficient to measure \Vtan\ for the
slower moving objects in the sample. Note that the 30\,\kms\ rms we
report for our sample includes scatter due to measurement error, and
the mean error in \Vtan\ for Y0~dwarfs is 9\,\kms. In computing the
mean and rms we considered only free-floating late-type systems, i.e.,
excluding companions to stars or other brown dwarfs, like Ross~458C
and WISEP~J1217+1626B. We also exclude three Y0--Y1 dwarfs that only
have parallaxes and proper motions from \cite{2013ApJ...762..119M},
since those authors used the \Vtan\ distribution of T~dwarfs as a
prior in their Bayesian astrometric analysis, thus their \Vtan\ values
are technically not independent of that prior.

We have checked if any objects in our sample could be likely members
of the thick disk or halo populations on the basis of having very high
\Vtan. Using criteria provided by \cite{2012ApJS..201...19D} for
determining if objects are likely non-thin disk members, we find that
none of our sample fit the criterion for $p_{\rm thin} < 0.1$. The two
objects with the most significant high \Vtan\ measurements are
PSO~J043.5+02.39 (T8; $91^{+12}_{-10}$\,\kms) and WISEP~J0410+1502
(Y0; $87^{+12}_{-10}$\,\kms), which are the only objects that do not
satisfy the $p_{\rm thin} > 0.9$ thin disk criterion from
\cite{2012ApJS..201...19D}. The only object with nominally higher
\Vtan\ than these two objects is WISEP~J0148$-$7202 (T9.5;
$99^{+40}_{-23}$\,\kms), but its distance is currently very uncertain
(${\rm S/N} = 3.8$).

\noindent \underline{Beyond Y0}

There are four objects with distance measurements that are classified
as having spectral types later than Y0. Two of these are in our
\Spitzer\ sample, WISEP~J1541$-$2250 (Y0.5) and WISEP~J1828+2650
($\geq$Y2), and the other two only have parallaxes from
\cite{2013ApJ...762..119M}, WISE~J0350$-$5658 (Y1) and
WISE~J0535$-$7500 ($\geq$Y1). These latter two have parallax S/N of
5.8 and 3.2, respectively, but \cite{2013ApJ...762..119M} report
distances having 1.6--1.8$\times$ lower S/N after applying their
Bayesian priors. This makes the distance of WISE~J0535$-$7500
particularly uncertain, since its parallax of $250\pm79$\,mas implies
a distance of 4\,pc but its quoted final distance is
$21^{+13}_{-11}$\,pc after applying priors \cite{2013ApJ...762..119M}.
Therefore we focus on the two objects with more robust distances in
the following.

WISEP~J1541$-$2250 (Y0.5) was originally reported as a Y0~dwarf at
2.8\,pc based on a preliminary parallax of $350\pm110$\,mas
\cite{2011ApJS..197...19K}. In previous work it has thus often
appeared as an extremely faint data point $\approx$5--6\,mag below the
end of the T~dwarf sequence. The same team has since revised its
spectral type to Y0.5 \cite{2012ApJ...753..156K} and parallax to
$-21\pm94$\,mas \cite{2013ApJ...762..119M}. Our parallax of
$74\pm31$\,mas is the most precise yet but still too low S/N to
securely determine its location relative to T8--Y0 dwarfs. It now
appears to be only $\approx$2\,mag fainter than the end of the
T~sequence in the near-infrared, and its absolute magnitude is consistent
with Y0~dwarfs in the near-infrared and with all T8--Y0 dwarfs in the
mid-infrared. Its $YJH$ colors also appear to be consistent with other
Y0~dwarfs, while its $[3.6]-[4.5]$ color is slightly redder than the
reddest Y0, the peculiar WISEP~J1405+5534.

WISE~J0350$-$5658 is defined as the spectral standard for the Y1 class
\cite{2012ApJ...753..156K}. This object does not have $YJHK$
photometry, but on mid-infrared color--magnitude diagrams it is the
reddest known object, with $[3.6]$ and $[4.5]$ absolute magnitudes
$\approx$2--3\,mag and $\approx$1--2\,mag fainter (1$\sigma$ ranges),
respectively, compared to Y0~dwarfs. Thus, it would appear that while
\Lbol\ does not drop substantially going from Y0 to Y0.5, it plummets
going from Y0.5 to Y1. Unfortunately, these Y0.5--Y1 objects have some
of the lowest significance distance measurements, and the sample is
very small, so it is unclear if these trends will turn out to be real.
At face value, WISE~J0350$-$5658 (Y1) would be the least luminous and
thus coldest (230--300\,K) object in the entire sample, aside from two
objects with very uncertain parallaxes, WISE~J1639$-$6847 and
WISE~J0359$-$5401. Therefore, it is remarkable that the near-infrared
spectrum of WISE~J0350$-$5658 \cite{2012ApJ...753..156K} is only
subtly different from other Y~dwarfs that are nominally
$\approx$150\,K (1.6$\times$) warmer.

WISEP~J1828+2650 has the highest precision distance of any object
later than Y0 ($14.3^{+3.6}_{-2.4}$\,pc). It was originally typed
as $>$Y0 \cite{2011ApJ...743...50C} but has been re-classified as
$\geq$Y2 \cite{2012ApJ...753..156K} in the context of a larger
sample of near-infrared spectra. WISEP~J1828+2650 has singular properties
not observed in any other Y~dwarf, and thus it has been the subject of
extensive discussion in the literature.
WISEP~J1828+2650 has been dubbed the archetype for the Y~spectral
class in prior work that ascribes its unusually red $J-H$ and
near-infrared minus mid-infrared colors to the collapse of flux as the
Wien tail moves into the near-infrared \cite{2011ApJ...743...50C}.
WISEP~J1828+2650 is also the only known object to show a suppressed
$J$-band, i.e., 1.27\,$\mu$m, flux peak. These authors estimated an
upper limit of $\Teff \lesssim 300$\,K based on a comparison of the
observed properties to predictions from atmospheric models
\cite{2011ApJ...743...50C}.
In subsequent work, a preliminary parallax of $122\pm13$\,mas
for WISEP~J1828+2650 to compute its absolute magnitudes
\cite{2012ApJ...753..156K}. This parallax is 3.4$\sigma$ larger than
the final value determined by \cite{2013ApJ...764..101B} and
2.7$\sigma$ larger than our \Spitzer-only parallax. Even using the
closer distance, \cite{2012ApJ...753..156K} found that
WISEP~J1828+2650 had similar or brighter magnitudes than Y1~dwarfs in
the near-infrared and surprisingly was as bright as Y0~dwarfs in the
mid-infrared. They suggested that if this rebounding of the flux
relative to earlier type objects is not a real effect, then it may be
due either to systematic errors in their preliminary parallaxes,
misclassification of the type for WISEP~J1828+2650, or some unknown
physical cause. Our updated parallax places WISEP~J1828+2650 even
farther away, intensifying this puzzle as its mid-infrared magnitudes
are even brighter than the earlier type Y~dwarfs.
Other authors have pointed out that for such large amount of flux to
be produced by an object with $\Teff \lesssim 300$\,K would require an
unusually large radius, implying a very young age of $\lesssim$50\,Myr
and low mass of $\lesssim$1\,\Mjup\ \cite{2013ApJ...763..130L}. They
suggested instead that WISEP~J1828+2650 is an unresolved binary
composed of 300\,K and 325\,K components with types of Y1 and Y1.5
\cite{2013ApJ...763..130L}. While this helps explain its mid-infrared
flux somewhat, bringing it into better agreement with models, it does
not explain the unusually red $J-H$ color and unique near-infrared
spectral morphology.
Recent work fitting the $H$- and $[4.5]$-band absolute magnitudes of
WISEP~J1828+2650 to Cond model isochrones has found $\Teff =
275\pm40$\,K and $\Teff = 450\pm40$\,K, respectively
\cite{2013ApJ...764..101B}. These authors conclude that the nature of
WISEP~J1828+2650 is currently ambiguous due to the fact that no models
consistently reproduce its properties \cite{2013ApJ...764..101B}.  It
has also been suggested that low metallicity may be partly responsible
for some of the unusual properties of WISEP~J1828+2650
\cite{2013MNRAS.433..457B}.

We find that WISEP~J1828+2650, with $\log(\Lbol/\Lsun) =
-6.13^{+0.20}_{-0.16}$\,dex, is at least as luminous as normal
Y0~dwarfs, which have a mean $\log(\Lbol/\Lsun)$ of $-$6.52\,dex. Our
approach of summing the near-infrared and mid-infrared fluxes to
estimate the bolometric luminosity is distinct from prior work, since
it greatly reduces the dependence of our results on predicted colors
and magnitudes from models. WISEP~J1828+2650 cannot be much less
luminous than we have estimated since we directly account for
$(1.3^{+0.8}_{-0.4})\times10^{27}$\,erg\,s$^{-1}$ of its luminosity
from photometry alone, which is equivalent to $\log(\Lbol/\Lsun) =
-6.45^{+0.19}_{-0.16}$\,dex. If WISEP~J1828+2650 is indeed cooler than
other Y~dwarfs, it must have a very large radius. To agree with the
300\,K upper limit in \Teff\ proposed by \cite{2011ApJ...743...50C}
would require $\Rstar = 3.1^{+0.8}_{-0.5}$\,\Rjup (or
$0.32^{+0.08}_{-0.05}$\,\Rsun), which is 3.5$\sigma$ larger than Cond
model radii even for an age of 10\,Myr. We therefore suggest that such
an explanation is unphysical and that the temperature of
WISEP~J1828+2650 is indeed warmer than expected from comparison to
model spectra. Our calculations give $\Teff = 520^{+60}_{-50}$\,K and
$560^{+80}_{-60}$\,K at ages of 1\,Gyr and 5\,Gyr, respectively. Even
if we split the flux into two hypothetical binary components as
proposed in Table~7 of \cite{2013ApJ...763..130L}, we find $\Teff =
440^{+60}_{-40}$\,K and $420^{+60}_{-40}$\,K at 1\,Gyr and $\Teff =
470^{+70}_{-50}$\,K and $450^{+60}_{-40}$\,K at 5\,Gyr. Thus,
regardless of whether or not it is an unresolved binary, we find that
WISEP~J1828+2650 is as warm or warmer than Y0~dwarfs. We note that it
could still be slightly younger than the Y0~dwarfs, and indeed models
show that the collapse of the near-infrared flux due to the Wien tail
happens at warmer \Teff\ when the surface gravity is lower even up to
ages of 5\,Gyr \cite{2003ApJ...596..587B}. 

\noindent \underline{Notes on Individual Objects}

We now discuss the handful of objects that are thought to be
spectrally peculiar, are lacking spectral types, or that otherwise
stand out as atypical.
\begin{itemize}

\item 2MASSW~J1225$-$2739B (T8): This is the brightest T8~dwarf in our
  sample by $\approx$0.6\,mag in the near-infrared.  If this object is
  an unresolved binary it would make this system a triple.
  However, we note that its T8 spectral type was not determined using
  resolved spectroscopy of the individual components, like most other
  objects in our sample but rather by a spectral decomposition
  technique matching summed template spectra to its integrated-light
  spectrum, constrained by the measured near-infrared flux ratios
  which has a quoted uncertainty of $\pm$0.5 subtypes
  \cite{2012ApJS..201...19D}.  This is consistent with previous
  estimates of T$8\pm1$ based on optical colors
  \cite{2003ApJ...586..512B} and T$7.5\pm0.5$ from near-infrared
  magnitudes \cite{2010ApJ...722..311L}. Thus, a simpler solution to
  its apparent overluminousness would be that its spectral type is
  T7.5, not T8, since this would be make its near-infrared magnitudes
  agree very well with the mean properties of other normal T7.5 dwarfs
  \cite{2012ApJS..201...19D}.

\item UGPS~J0722$-$0540 (T9): This object is defined as the spectral
  standard for the T9 class  \cite{2011ApJ...743...50C}.  However,
  its location on color--magnitude diagrams is notably distinct from
  the other three T9~dwarfs (Fig.~\ref{fig:cmd}); its absolute
  magnitudes are $\approx$0.8\,mag fainter in the near-infrared and
  $\approx$0.6\,mag fainter in the mid-infrared, and its derived $\Teff
  \approx 500$--550\,K is correspondingly $\approx$100\,K lower.  As
  more parallaxes for T9~dwarfs are obtained, it should become clear
  whether UGPS~J0722$-$0540 is indeed unique or if the T9 subclass
  simply shows an unusual amount of diversity.

\item WISEP~J1405+5534 (Y0p): This object has similar mid-infrared
  absolute magnitudes to normal Y0~dwarfs, but our parallaxes show
  that it is $\approx$1.5\,mag fainter in the near-infrared.  Thus, we
  find it has essentially the same luminosity and temperature as other
  Y0~dwarfs, but different underlying properties.  At discovery this
  object was typed as peculiar (``pec?'') due to the fact that its
  spectrum's $H$-band peak is shifted 60\,\AA\ redder than the Y0
  standard WISEP~J1738+2732 \cite{2011ApJ...743...50C}. If this shift
  is due to enhanced NH$_3$ absorption as compared to other Y0~dwarfs
  of the same \Teff, we suggest this may imply a reduced level of
  nonequilibrium chemistry in the photosphere, perhaps due to reduced
  vertical mixing.  This would be also be consistent with the fact
  that WISEP~J1405+5534 is the reddest Y0~dwarf in $[3.6]-[4.5]$,
  which implies enhanced CH$_4$ absorption and reduced CO absorption.
  However, it is not clear how this would lead to a $\approx$1.5\,mag
  flux drop in the $YJHK$ bands but no difference in the mid-infrared
  flux.

\item WISE~J0359$-$5401 (Y0): This object only has a parallax from
  \cite{2013ApJ...762..119M}, who find $145\pm39$\,mas and $d =
  5.9^{+1.3}_{-0.8}$\,pc after applying their Bayesian priors.
  However, we find it implausible that the precision in the distance
  (S/N = 5--7) could actually be higher than the parallax (S/N = 3.7),
  and thus we have excluded this object from our analysis until a more
  precise parallax is available. (At the quoted distance from
  \cite{2013ApJ...762..119M}, WISE~J0359$-$5401 would be the faintest
  known Y~dwarf, 2--3\,mag fainter in both near-infrared and
  mid-infrared bands as compared to other Y0~dwarfs.)

\item WISE~J1639$-$6847 (Y0:): This object has a preliminary parallax
  of 200\,mas based on three epochs spanning three months in 2012
  using the near-infrared camera FourStar at Magellan combined with
  two \WISE\ epochs \cite{2012ApJ...759...60T}.  The formal parallax
  uncertainty is 12\,mas, and these authors adopt an error of 20\,mas
  \cite{2012ApJ...759...60T}.  This object only has $J$-band
  photometry, and the preliminary distance implies that it would be
  $\approx$2\,mag fainter than normal Y0~dwarfs. Using the same
  rationale as for WISE~J0359$-$5401 above, we have excluded this
  object from our analysis until a parallax is measured with longer
  time baseline data.

\item CFBDSIR~J1458+1013B: This object was discovered by
  \cite{2011ApJ...740..108L}, who speculated that its low luminosity
  and unusual near-infrared colors might ultimately lead it to be classified
  as a Y~dwarf.  A spectrum for this 0$\farcs$11 binary companion is
  not available; however, we can now compare its near-infrared absolute
  magnitudes to those of other Y~dwarfs.  On various $YJHK$
  color--magnitude diagrams, it appears to be slightly brighter than
  or consistent with Y0~dwarfs.  CFBDSIR~J1458+1013B is
  $\approx$1\,mag fainter than the faintest T9~dwarf and 0.4--0.7\,mag
  brighter than the mean of normal Y0~dwarfs.  Thus, we find that it
  is more likely to be an early Y~dwarf than a very late T~dwarf, and
  we suggest using a photometric spectral type estimate of Y0.

\item WD~0806$-$661B: This object has only been detected in two
  bandpasses, $[3.6]$ and $[4.5]$. According to our estimate of
  $\log(\Lbol/\Lsun) = -6.81\pm0.09$\,dex, this would be the least
  luminous object known to date, possibly except for a few objects
  that have very uncertain distances (WISE~J0350$-$5658,
  WISE~J0359$-$5401, and WISE~J1639$-$6847). Using the white dwarf
  cooling age from WD~0806$-$661A and Cond models we find $\Teff =
  353^{+23}_{-22}$\,K, slightly higher than the 300--345\,K suggested
  by \cite{2012ApJ...744..135L} based on their comparison of the
  $[4.5]$-band absolute magnitude and $J$-band nondetection to models
  of \cite{2003ApJ...596..587B,2008ApJ...689.1327S}.  As these
  authors point out, the models do not reproduce the location of
  WD~0806$-$661B on mid-infrared color--magnitude diagrams
  \cite{2012ApJ...744..135L}, so using a single-band flux measurement
  for their \Teff\ is likely less robust than using the combined flux
  in both \Spitzer\ bandpasses as we have done. As discussed above,
  the case of WISEP~J1828+2650 also highlights the perils of inferring
  fundamental properties from model predicted colors and magnitudes
  for such cool objects.

\end{itemize}

\clearpage

\begin{landscape}

\begin{figure}

\centerline{
\includegraphics[width=2.1in,angle=0]{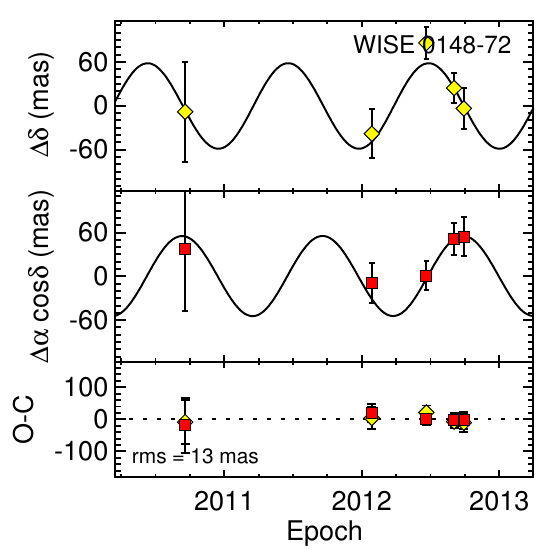}
\includegraphics[width=2.1in,angle=0]{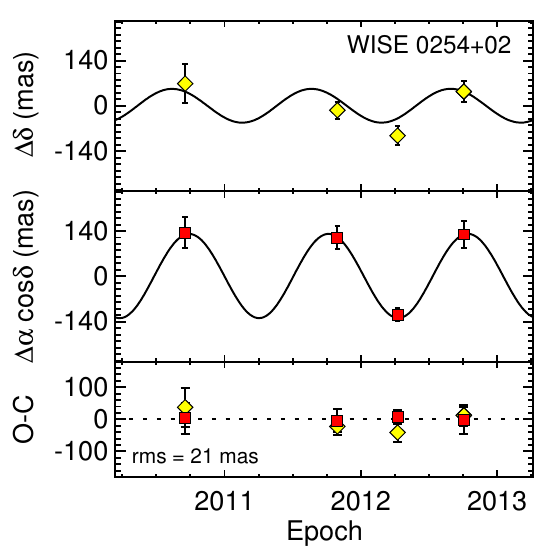}
\includegraphics[width=2.1in,angle=0]{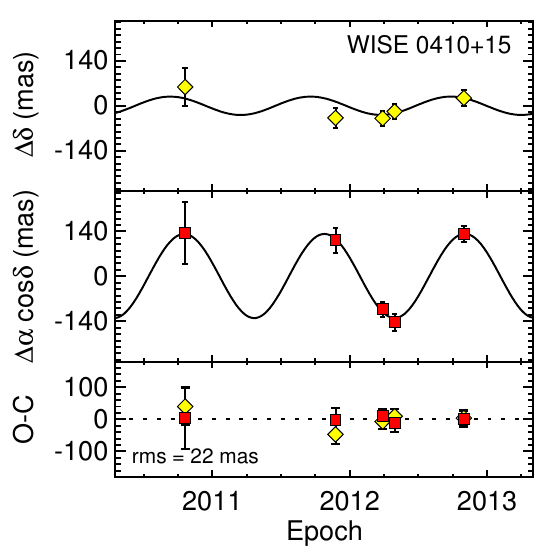}
\includegraphics[width=2.1in,angle=0]{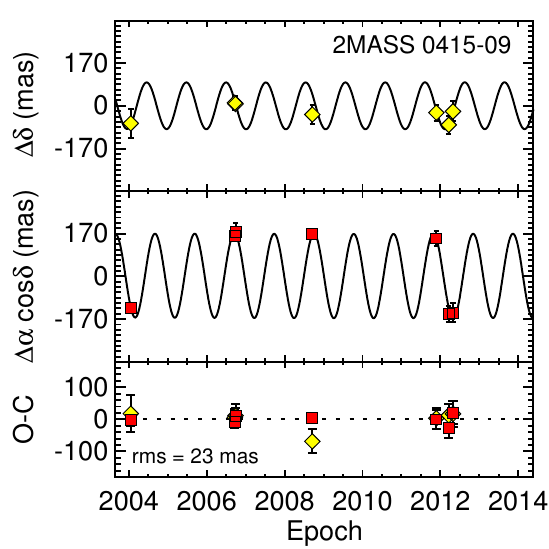}
}
\centerline{
\includegraphics[width=2.1in,angle=0]{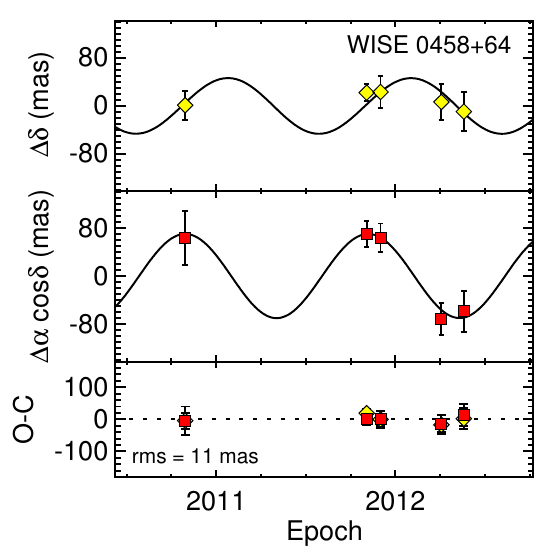}
\includegraphics[width=2.1in,angle=0]{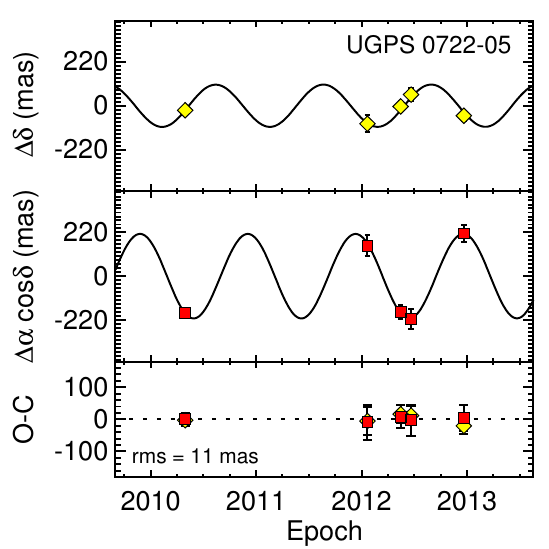}
\includegraphics[width=2.1in,angle=0]{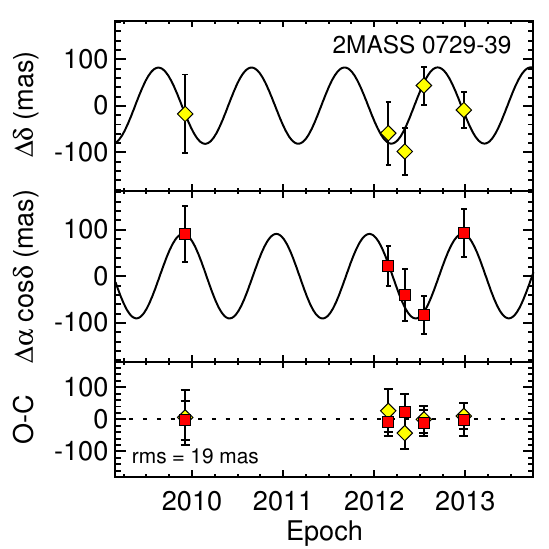}
\includegraphics[width=2.1in,angle=0]{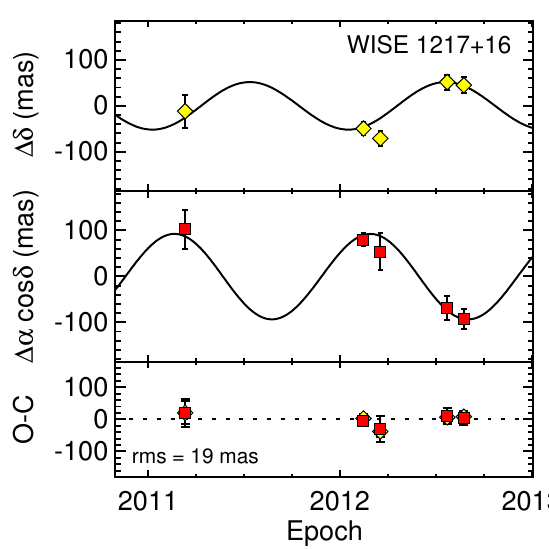}
}

\caption{\normalsize For each object, the top and middle panels show
  relative astrometry in $\delta$ and $\alpha$, respectively, as a
  function of Julian year after subtracting the best-fit proper
  motion.  (This is for display purposes only; in our analysis we fit
  for both the proper motion and parallax simultaneously.)  The bottom
  panels show the residuals after subtracting both the parallax and
  proper motion and give the rms of the data about the fit. \label{fig:plx}}

\end{figure}

\begin{figure}

\centerline{
\includegraphics[width=2.1in,angle=0]{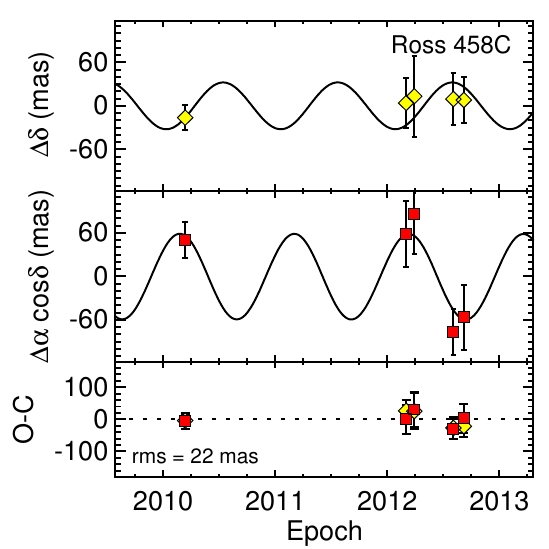}
\includegraphics[width=2.1in,angle=0]{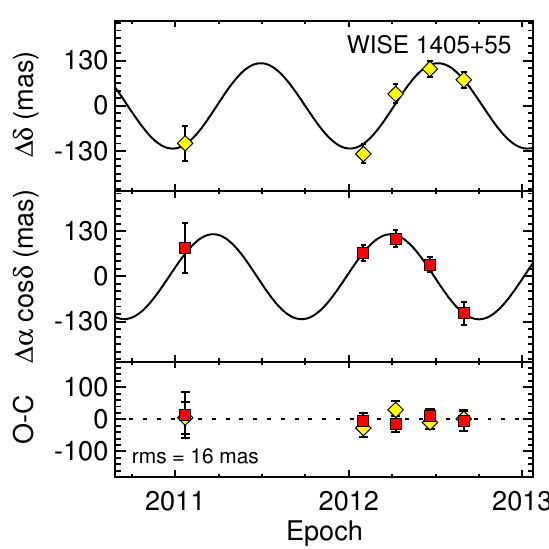}
\includegraphics[width=2.1in,angle=0]{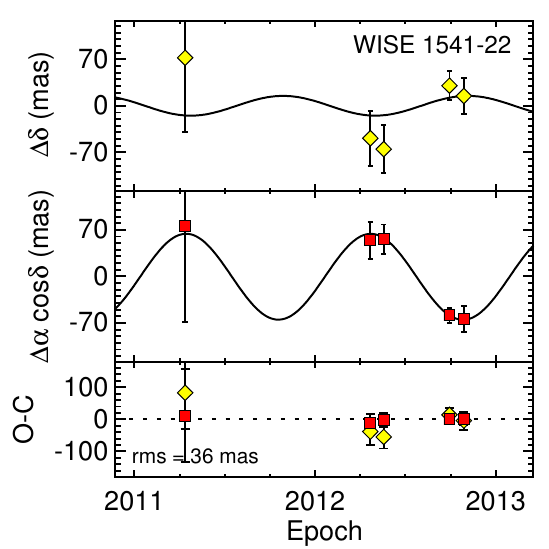}
\includegraphics[width=2.1in,angle=0]{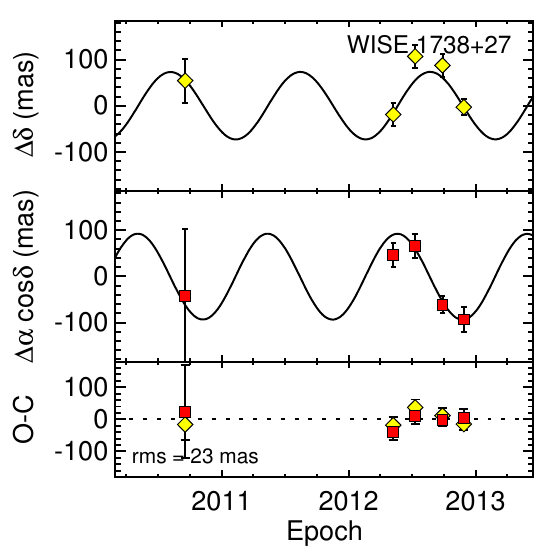}
}
\centerline{
\includegraphics[width=2.1in,angle=0]{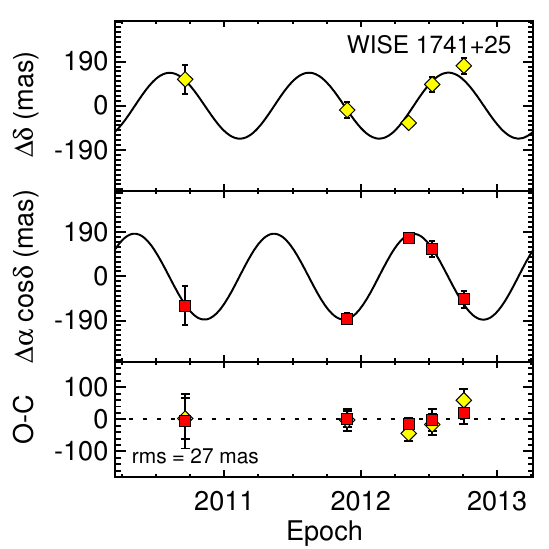}
\includegraphics[width=2.1in,angle=0]{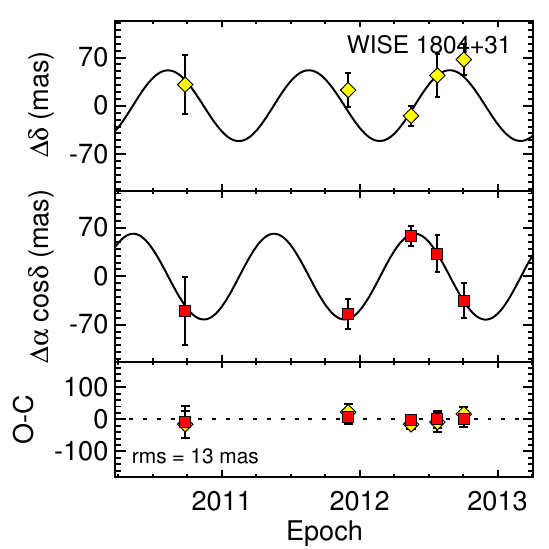}
\includegraphics[width=2.1in,angle=0]{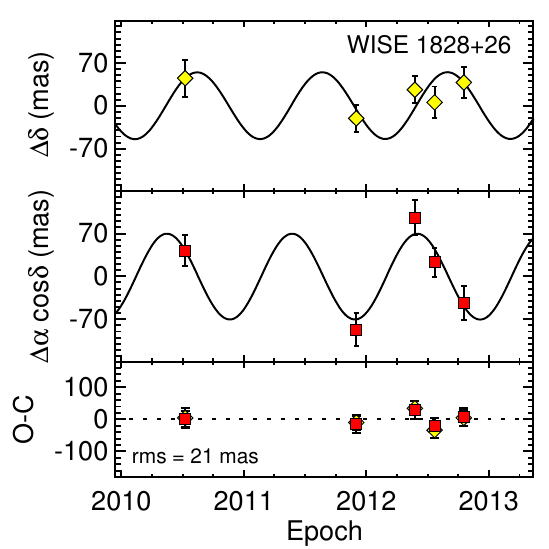}
\includegraphics[width=2.1in,angle=0]{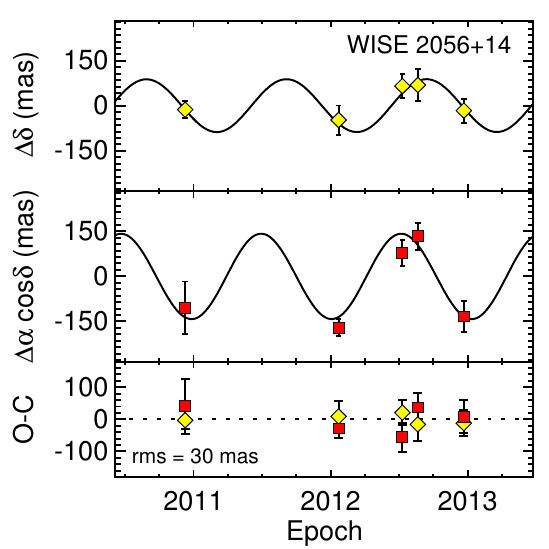}
}

\ContinuedFloat
\caption{\normalsize (Continued) \label{fig:plx}}

\end{figure}

\end{landscape}
\clearpage

\begin{figure}

\centerline{\includegraphics[width=3.1in,angle=0]{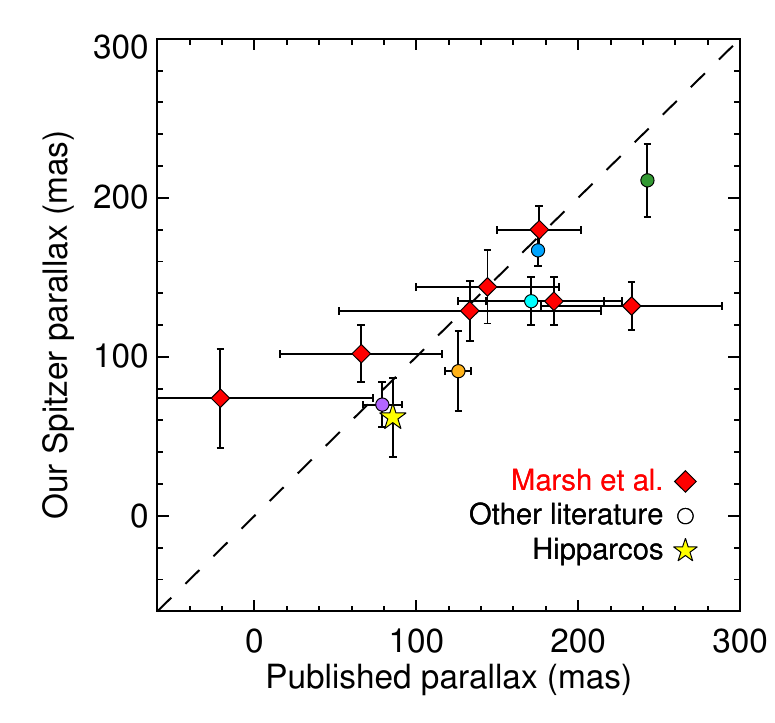}}

\caption{\normalsize Comparison of our \Spitzer\ parallaxes to
  published values. The largest single comparison sample are seven of
  our science targets that also have parallaxes from
  \cite{2013ApJ...762..119M}, shown as red
  diamonds. We also observed a control sample of
  five late-T dwarfs that had previously published parallaxes. The
  $\chi^2$ computed from the differences between our parallaxes and
  published values is reasonable for both our control sample ($\chi^2
  = 5.7$, 5~dof) and the Marsh et~al.\ sample
  ($\chi^2 = 5.7$, 7~dof). \label{fig:vs-pub}}


\end{figure}

\clearpage

\begin{figure}

\centerline{\includegraphics[width=3.1in,angle=0]{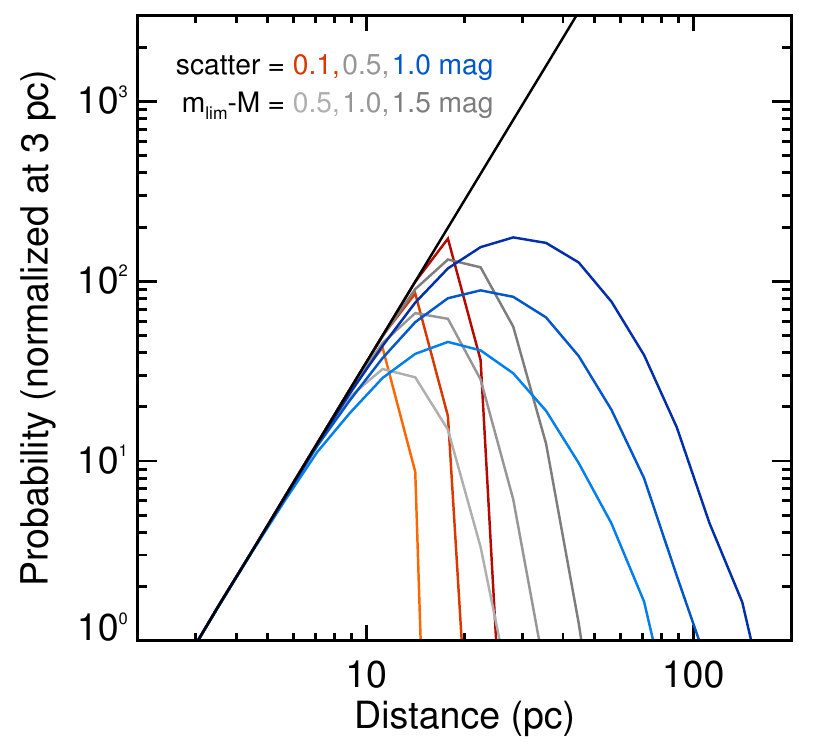}}

\caption{\normalsize Distance priors assuming objects are uniformly
  distributed in volume (straight black line) or discovered in a
  magnitude limited survey (colored/gray lines).  These simulations
  assume a single class of objects that have a mean absolute magnitude
  $M$ and intrinsic scatter of 0.1\,mag (red/orange), 0.5\,mag (gray),
  or 1.0\,mag (blue).  We consider limiting magnitudes ($m_{\rm lim}$)
  that are 0.5\,mag, 1.0\,mag, or 1.5\,mag fainter than the mean
  absolute magnitude of the objects, and the resulting distance
  distributions are normalized to unity at 3\,pc.  This shows that a
  prior uniform in volume is not appropriate for the vast majority of
  objects discovered in a magnitude limited survey as only the very
  nearest objects follow such a distribution.  Most objects would be
  discovered near the peaks of these probability distributions, and
  thus a flat prior in distance and parallax would be more reasonable.
  Since we do not know the absolute magnitudes of Y~dwarfs a priori,
  we conservatively assume a flat prior in parallax in our astrometric
  analysis. \label{fig:lutz}}

\end{figure}

\clearpage

\begin{figure}

\centerline{\includegraphics[width=5.0in,angle=0]{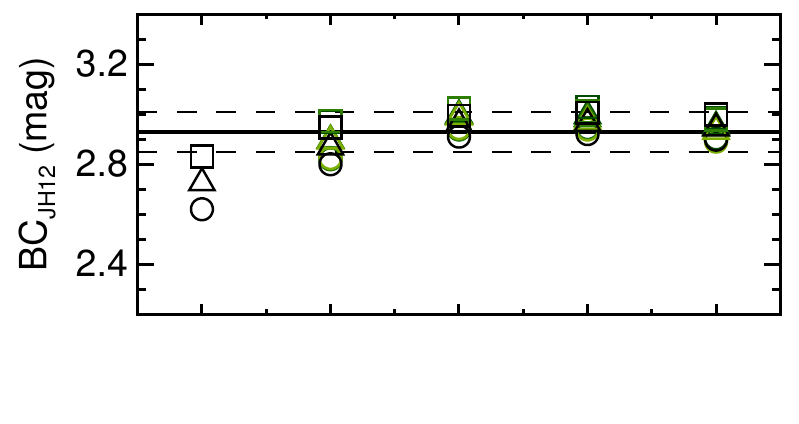}}
\vskip -0.70in
\centerline{\includegraphics[width=5.0in,angle=0]{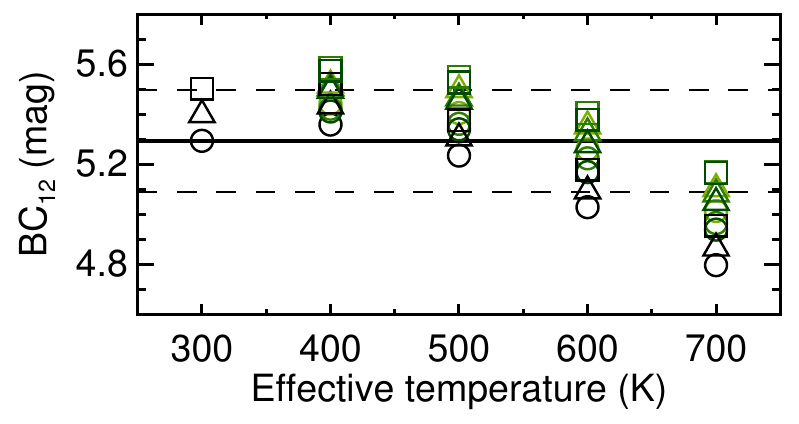}}

\caption{\normalsize Bolometric corrections derived from model
  atmospheres \cite{2012ApJ...756..172M,2012ApJ...750...74S} for the
  super-magnitudes $m_{JH12}$ (top) and $m_{12}$ (bottom).  The mean
  for each set of models is shown as a solid line, with dashed lines
  showing $\pm$1$\sigma$ uncertainty from the rms.  We used models
  with surface gravities of $10^4$\,cm\,s$^{-2}$ (circles), $3 \times
  10^4$\,cm\,s$^{-2}$ (triangles), and $10^5$\,cm\,s$^{-2}$ (squares).
  The shades of symbols indicate either cloud-free models (black) or
  $f_{\rm sed} = 3$, 4, or 5 (light, medium, and dark green,
  respectively), with larger $f_{\rm sed}$ corresponding to thinner
  clouds.  The very weak dependence of these bolometric corrections on
  effective temperature enables us to adopt a single value for each
  super-magnitude for our entire sample of late-T and
  Y~dwarfs. \label{fig:bc}}

\end{figure}

\clearpage

\begin{figure}

\centerline{\includegraphics[width=5.0in,angle=0]{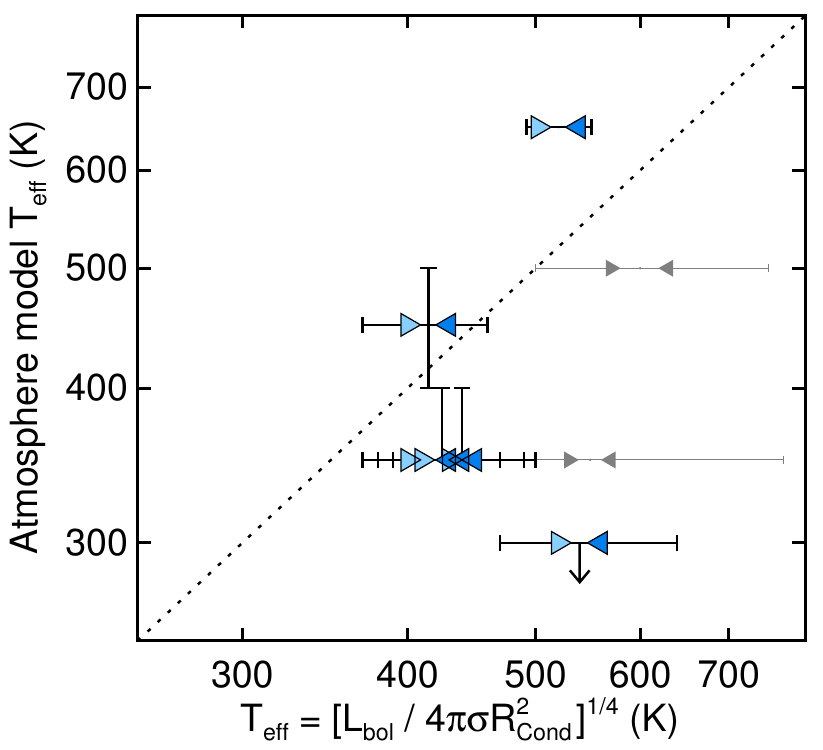}}

\caption{\normalsize Effective temperatures estimated from model
  atmosphere fitting \cite{2011ApJ...743...50C} plotted as a function
  of our luminosity-based \Teff\ estimates.  Symbols are the same as
  in Fig.~\ref{fig:lbol-teff}, with triangles showing the \Teff\ at
  ages of 1\,Gyr and 5\,Gyr and $x$-axis error bars indicating the
  uncertainty due to luminosity errors.  Objects with very uncertain
  \Lbol\ are shown in gray.  We plot the best-fit model atmosphere
  \Teff\ with error bars showing the range of model parameters
  consistent with the data reported in Table~6 of
  \cite{2011ApJ...743...50C}.  We find that the Y~dwarfs with model
  atmosphere temperatures of 350\,K are in fact significantly warmer
  ($\approx$400--500\,K).  The most extreme case is WISEP~J1828+2650
  ($\geq$Y2) for which model atmospheres give an upper
  limit of $\leq$300\,K \cite{2011ApJ...743...50C}, and we find $\approx$450--650\,K.
  WISEP~J0410+1502 is the only Y~dwarf for which model atmospheres
  predict a \Teff\ in agreement with our values, but this is also the
  lowest gravity model atmosphere fit ($\logg = 3.75$) that would
  imply a young age and mass of only 3\,\Mjup.  UGPS~J0722$-$0540 (T9)
  has a best-fit model atmosphere \Teff\ of 650\,K, much warmer than
  our luminosity-based estimate of $\approx$500--550\,K.  \label{fig:teff-teff}}

\end{figure}

\clearpage


\section*{Supplementary material}

Table S1. Parallax and Proper Motion MCMC Results

| Target | $\pi_{\rm rel}$ ($''$) | $\alpha_{J2000}$ (deg) | $\delta_{J2000}$ (deg) | Epoch (MJD) | $\mu_\alpha \cos\delta$ ($''$ year$^{-1}$) | $\mu_\delta$ ($''$ year$^{-1}$) | $\mu$ ($''$ year$^{-1}$) | P.A. (deg) | $\chi^2$/dof | $N_{\rm ep}$ | $\Delta t$ (yr) | $N_{\rm ref}$ | $N_{\rm cal}$ |
|---|---|---|---|---|---|---|---|---|---|---|---|---|---|
| WISEP J014807.25−720258.8 | 0.060(16) | 027.03049 | −72.04965 | 55458.49 | 1.257(38) | −0.008(33) | 1.257(38) | 90.4 ± 1.5 | 1.9/5 | 5 | 2.03 | 207 | 27 |
| PSO J043.5395+02.3995[a] | 0.135(15) | 043.53988 | +02.39964 | 55456.99 | 2.588(27) | 0.273(27) | 2.602(27) | 84.0 ± 0.6 | 3.3/3 | 4 | 2.04 | 100 | 19 |
| WISEP J041022.71+150248.5 | 0.132(15) | 062.59494 | +15.04641 | 55490.07 | 0.958(37) | −2.229(29) | 2.426(30) | 156.7 ± 0.9 | 3.4/5 | 5 | 2.03 | 154 | 19 |
| 2MASSI J0415195−093506 | 0.167(10) | 063.83469 | −09.58437 | 53024.06 | 2.204(3) | 0.541(46) | 2.269(3) | 76.21 ± 0.11 | 5.5/9 | 7 | 8.27 | 75 | 17 |
| WISEP J045853.90+643451.9 | 0.070(19) | 074.72478 | +64.58131 | 55498.85 | 0.136(45) | 0.317(22) | 0.347(26) | 23 ± 7 | 2.7/5 | 5 | 1.56 | 124 | 28 |
| UGPS J072227.51−054031.2 | 0.211(23) | 110.61371 | −05.67496 | 55316.49 | −0.893(16) | 0.349(10) | 0.959(15) | 291.3 ± 0.7 | 1.3/5 | 5 | 2.64 | 84 | 46 |
| 2MASS J07290002−3954043 | 0.091(25) | 112.24794 | −39.89626 | 55169.61 | −0.540(25) | 1.694(30) | 1.778(30) | 342.3 ± 0.8 | 1.2/5 | 5 | 3.06 | 105 | 60 |
| WISEP J121756.91+162640.2 | 0.099(16) | 184.48726 | +16.44439 | 55632.99 | 0.786(42) | −1.224(27) | 1.455(38) | 147.3 ± 1.3 | 7.0/5 | 5 | 1.45 | 103 | 8 |
| Ross 458C | 0.062(25) | 195.17351 | +12.35406 | 55268.87 | −0.649(20) | −0.045(12) | 0.650(20) | 266.1 ± 1.1 | 3.2/5 | 5 | 2.49 | 62 | 11 |
| WISEP J140518.40+553421.5 | 0.129(19) | 211.32589 | +55.57258 | 55583.18 | −2.263(47) | 0.288(41) | 2.281(48) | 277.3 ± 1.0 | 3.0/5 | 5 | 1.61 | 131 | 17 |
| WISEP J154151.65−225025.2 | 0.074(31) | 235.46470 | −22.84053 | 55664.92 | −0.870(130) | −0.013(58) | 0.870(130) | 269 ± 4 | 4.3/5 | 5 | 1.54 | 216 | 40 |
| WISEP J173835.52+273258.9 | 0.102(18) | 264.64812 | +27.54967 | 55457.58 | 0.292(63) | −0.396(22) | 0.493(40) | 144 ± 6 | 6.4/5 | 5 | 2.19 | 121 | 22 |
| WISEP J174124.26+255319.5 | 0.180(15) | 265.35094 | +25.88859 | 55457.57 | −0.509(35) | −1.463(32) | 1.550(33) | 199.2 ± 1.3 | 7.9/5 | 5 | 2.04 | 61 | 21 |
| WISEP J180435.40+311706.1 | 0.060(11) | 271.14730 | +31.28515 | 55465.21 | −0.242(26) | 0.017(22) | 0.244(26) | 274 ± 5 | 2.8/5 | 5 | 2.02 | 194 | 42 |
| WISEP J182831.08+265037.8 | 0.070(14) | 277.12961 | +26.84387 | 55387.33 | 1.020(15) | 0.173(16) | 1.034(15) | 80.4 ± 0.9 | 6.5/5 | 5 | 2.28 | 209 | 45 |
| WISEP J205628.90+145953.3 | 0.144(23) | 314.12055 | +14.99824 | 55540.03 | 0.761(46) | 0.500(21) | 0.911(41) | 56.7 ± 1.9 | 4.0/5 | 5 | 2.03 | 74 | 30 |

[a] WISEP J025409.45+022359.1

Note. — This table gives all the astrometric parameters derived from our MCMC analysis. For parameters in units of arcseconds, errors are given in parentheses in units of milliarcsec. $\pi_{\rm rel}$: The parallax relative to the reference star frame. Corrections to the absolute frame are negligible here. ($\alpha$, $\delta$, MJD): Coordinates that correspond to the epoch listed, which is the first epoch of *Spitzer* observations for that target. ($\mu_\alpha \cos\delta$, $\mu_\delta$, $\mu$, P.A.): Proper motion parameters are listed both as the direct fitting results (i.e., in $\alpha$ and $\delta$) and the computed quantities of total amplitude ($\mu$) and position angle. $\chi^2$/dof: The lowest $\chi^2$ in each set of MCMC chains along with the degrees of freedom. $N_{\rm ref}$: Total number of reference stars used. $N_{\rm cal}$: Subset of reference stars from the *WISE* All-Sky Source Catalog used in the absolute astrometric calibration of constant and linear terms.

29

## Table S2. Measurements of Late-T and Y dwarfs with Parallaxes

| Object | Spec. Type | Distance (pc) | $Y$ (mag) | $J$ (mag) | $H$ (mag) | $K$ (mag) | [3.6] (mag) | [4.5] (mag) | References |
|---|---|---|---|---|---|---|---|---|---|
| PSO J043.5395+02.3995[a] | T8 | $7.4^{+0.9}_{-0.7}$ | ⋯ | $16.14 \pm 0.12$ | $16.51 \pm 0.12$ | $16.84 \pm 0.12$ | $14.67 \pm 0.02$ | $12.68 \pm 0.02$ | (0,33) |
| 2MASSI J0415195−093506 | T8 | $5.71^{+0.06}_{-0.05}$ | $16.28 \pm 0.03$ | $15.32 \pm 0.03$ | $15.70 \pm 0.03$ | $15.83 \pm 0.03$ | $14.10 \pm 0.04$ | $12.29 \pm 0.03$ | (12,54,59,64) |
| 2MASS J07290002−3954043 | T8p | $7.9 \pm 0.5$ | $16.52 \pm 0.08$ | $15.64 \pm 0.08$ | $16.05 \pm 0.18$ | ⋯ | $14.47 \pm 0.01$ | $12.95 \pm 0.01$ | (12,31,50,62) |
| 2MASS J09393548−2448279 | T8 | $5.35^{+0.15}_{-0.14}$ | $16.47 \pm 0.09$ | $15.61 \pm 0.09$ | $15.96 \pm 0.09$ | $16.83 \pm 0.09$ | $13.76 \pm 0.02$ | $11.66 \pm 0.02$ | (40,54,60) |
| 2MASSW J1225543−273947B | T8 | $13.3^{+0.5}_{-0.4}$ | ⋯ | $16.48 \pm 0.03$ | $16.91 \pm 0.03$ | $17.10 \pm 0.03$ | ⋯ | ⋯ | (12,66) |
| Ross 458C | T8 | $11.70^{+0.21}_{-0.20}$ | $17.54 \pm 0.02$ | $16.71 \pm 0.01$ | $17.07 \pm 0.03$ | $16.96 \pm 0.03$ | $15.28 \pm 0.01$ | $13.77 \pm 0.01$ | (3,34,57) |
| BD+01 2920B | T8p | $17.18 \pm 0.15$ | $19.69 \pm 0.05$ | $18.71 \pm 0.05$ | $19.14 \pm 0.20$ | $19.89 \pm 0.33$ | $16.77 \pm 0.03$ | $14.71 \pm 0.01$ | (34,47) |
| ULAS J003402.77−005206.7 | T8.5 | $14.56^{+0.30}_{-0.29}$ | $18.90 \pm 0.10$ | $18.15 \pm 0.03$ | $18.49 \pm 0.04$ | $18.48 \pm 0.05$ | $16.28 \pm 0.03$ | $14.49 \pm 0.03$ | (3,12,68) |
| CFBDS J005910.90−011401.3 | T8.5 | $9.69^{+0.20}_{-0.19}$ | $18.82 \pm 0.02$ | $18.06 \pm 0.03$ | $18.27 \pm 0.05$ | $18.69 \pm 0.05$ | $15.71 \pm 0.01$ | $13.66 \pm 0.01$ | (3,12,50,58) |
| WISEP J045853.90+643451.9A | T8.5 | $14^{+5}_{-3}$ | ⋯ | $17.50 \pm 0.07$ | $17.77 \pm 0.11$ | ⋯ | ⋯ | ⋯ | (0,55) |
| WISE J111838.70+312537.9 | T8.5 | $8.29 \pm 0.15$ | $19.18 \pm 0.12$ | $17.79 \pm 0.05$ | $18.15 \pm 0.06$ | ⋯ | $15.60 \pm 0.03$ | $13.37 \pm 0.02$ | (48) |
| ULAS J133553.45+113005.2 | T8.5 | $10.01 \pm 0.16$ | $18.81 \pm 0.04$ | $17.90 \pm 0.01$ | $18.25 \pm 0.01$ | $18.28 \pm 0.03$ | $15.95 \pm 0.03$ | $13.91 \pm 0.03$ | (3,12,56) |
| Wolf 940B | T8.5 | $11.9^{+0.6}_{-0.5}$ | $18.97 \pm 0.04$ | $18.18 \pm 0.03$ | $18.77 \pm 0.03$ | $18.97 \pm 0.06$ | $16.44 \pm 0.03$ | $14.43 \pm 0.03$ | (3,46,61,67) |
| UGPS J072227.51−054031.2 | T9 | $4.12 \pm 0.04$ | $17.37 \pm 0.02$ | $16.52 \pm 0.02$ | $16.90 \pm 0.02$ | $17.07 \pm 0.08$ | $14.28 \pm 0.05$ | $12.19 \pm 0.04$ | (3,32,63) |
| WISEP J121756.91+162640.2A | T9 | $10.1^{+1.9}_{-1.4}$ | $18.59 \pm 0.04$ | $17.98 \pm 0.02$ | $18.31 \pm 0.05$ | $18.94 \pm 0.04$ | ⋯ | ⋯ | (0,10) |
| CFBDSIR J145829+101343A | T9 | $31.9^{+2.8}_{-2.4}$ | $20.81 \pm 0.21$ | $19.83 \pm 0.02$ | $20.18 \pm 0.10$ | $20.63 \pm 0.24$ | ⋯ | ⋯ | (2,10,12) |
| WISEP J174124.26+255319.5 | T9 | $5.6^{+0.5}_{-0.4}$ | $17.23 \pm 0.02$ | $16.18 \pm 0.02$ | $16.31 \pm 0.04$ | $17.02 \pm 0.20$ | $14.43 \pm 0.02$ | $12.39 \pm 0.02$ | (0,7) |
| WISEP J014807.25−720258.8 | T9.5 | $17^{+6}_{-4}$ | ⋯ | $18.96 \pm 0.07$ | $19.22 \pm 0.04$ | ⋯ | $16.84 \pm 0.05$ | $14.65 \pm 0.02$ | (0,3,7) |
| WISEP J045853.90+643451.9B | T9.5 | $14^{+5}_{-3}$ | ⋯ | $18.48 \pm 0.07$ | $18.79 \pm 0.11$ | ⋯ | ⋯ | ⋯ | (0,55) |
| WISEP J180435.40+311706.1 | T9.5: | $16.7^{+3.7}_{-2.6}$ | ⋯ | $18.70 \pm 0.05$ | $19.21 \pm 0.11$ | ⋯ | $16.62 \pm 0.04$ | $14.60 \pm 0.02$ | (0,7) |
| WISE J035934.06−540154.6 | Y0 | $5.9^{+1.3}_{-0.8}$ | ⋯ | $21.56 \pm 0.24$ | $22.20 \pm 0.43$ | ⋯ | $17.55 \pm 0.07$ | $15.33 \pm 0.02$ | (8,35) |
| WISEP J041022.71+150248.5 | Y0 | $7.6^{+1.0}_{-0.8}$ | $19.61 \pm 0.04$ | $19.44 \pm 0.03$ | $20.02 \pm 0.05$ | $19.91 \pm 0.07$ | $16.56 \pm 0.01$ | $14.12 \pm 0.01$ | (0,3,9) |
| WISEP J121756.91+162640.2B | Y0 | $10.1^{+1.9}_{-1.4}$ | $20.26 \pm 0.04$ | $20.08 \pm 0.03$ | $20.51 \pm 0.06$ | $21.10 \pm 0.12$ | ⋯ | ⋯ | (0,10) |
| WISEP J140518.40+553421.5 | Y0p | $7.8^{+1.3}_{-1.0}$ | $21.24 \pm 0.10$ | $21.06 \pm 0.06$ | $21.41 \pm 0.08$ | ⋯ | $16.78 \pm 0.01$ | $14.02 \pm 0.01$ | (0,3,9) |
| WISE J163940.83−684738.6 | Y0: | $5.0^{+0.6}_{-0.5}$ | ⋯ | $20.76 \pm 0.08$ | ⋯ | ⋯ | ⋯ | ⋯ | (53) |
| WISEP J173835.52+273258.9 | Y0 | $9.8^{+2.1}_{-1.5}$ | $19.86 \pm 0.08$ | $20.05 \pm 0.09$ | $20.45 \pm 0.09$ | $20.58 \pm 0.10$ | $16.87 \pm 0.01$ | $14.42 \pm 0.01$ | (0,3,9) |
| WISEP J205628.90+145953.3 | Y0 | $6.9^{+1.3}_{-1.0}$ | $19.77 \pm 0.06$ | $19.43 \pm 0.04$ | $19.96 \pm 0.04$ | $20.01 \pm 0.06$ | $15.90 \pm 0.01$ | $13.89 \pm 0.01$ | (0,3,9) |
| WISEP J154151.65−225025.2 | Y0.5 | $14^{+10}_{-4}$ | $21.46 \pm 0.13$ | $21.12 \pm 0.06$ | $22.17 \pm 0.25$ | ⋯ | $16.92 \pm 0.02$ | $14.12 \pm 0.01$ | (0,8,9) |
| WISE J035000.32−565830.2 | Y1 | $3.7^{+1.6}_{-0.4}$ | ⋯ | ⋯ | ⋯ | ⋯ | $17.94 \pm 0.10$ | $14.69 \pm 0.02$ | (8,35) |
| WISE J053516.80−750024.9 | ≥Y1: | $21^{+13}_{-11}$ | ⋯ | ⋯ | ⋯ | ⋯ | $17.77 \pm 0.09$ | $15.01 \pm 0.02$ | (8,35) |
| WISEP J182831.08+265037.8 | ≥Y2 | $14.3^{+3.6}_{-2.4}$ | $23.03 \pm 0.17$ | $23.48 \pm 0.23$ | $22.85 \pm 0.24$ | $23.48 \pm 0.36$ | $16.84 \pm 0.01$ | $14.27 \pm 0.01$ | (0,8,9) |
| CFBDSIR J145829+101343B | ⋯ | $31.9^{+2.8}_{-2.4}$ | $22.36 \pm 0.24$ | $21.85 \pm 0.06$ | $22.51 \pm 0.16$ | $22.83 \pm 0.30$ | ⋯ | ⋯ | (10,12) |
| WD 0806−661B | ⋯ | $19.2 \pm 0.6$ | ⋯ | ⋯ | ⋯ | ⋯ | $19.65 \pm 0.15$ | $16.88 \pm 0.05$ | (25,65) |

[a] WISEP J025409.45+022359.1

Note. — Compilation of distances, spectral types, and apparent magnitudes for all objects of spectral type T8 or later with parallax measurements. Uncertainties in spectral types are 0.5 subtypes unless otherwise noted (±1 subtype errors are denoted by ":"). All near-IR photometry is on the MKO system. Reference numbers correspond to the citations in the main article, with (0) refering to the work presented here.

Table S3. Mean Absolute Magnitudes of Normal Late-T and Y Dwarfs

| Spec. Type | $Y$ avg | $\sigma_{avg}$ | $\sigma_{add}$ | $n$ | $J$ avg | $\sigma_{avg}$ | $\sigma_{add}$ | $n$ | $H$ avg | $\sigma_{avg}$ | $\sigma_{add}$ | $n$ | $K$ avg | $\sigma_{avg}$ | $\sigma_{add}$ | $n$ | [3.6] avg | $\sigma_{avg}$ | $\sigma_{add}$ | $n$ | [4.5] avg | $\sigma_{avg}$ | $\sigma_{add}$ | $n$ |
|---|---|---|---|---|---|---|---|---|---|---|---|---|---|---|---|---|---|---|---|---|---|---|---|---|
| T8.0 | 17.40 | 0.03 | $\geq$0.25 | 3 | 16.43 | 0.02 | $\geq$0.46 | 5 | 16.82 | 0.03 | $\geq$0.43 | 5 | 16.93 | 0.03 | $\geq$0.80 | 5 | 15.11 | 0.03 | $\geq$0.15 | 4 | 13.40 | 0.02 | $\geq$0.21 | 4 |
| T8.5 | 18.81 | 0.03 | $\geq$0.51 | 5 | 17.87 | 0.02 | $\geq$0.44 | 6 | 18.20 | 0.03 | $\geq$0.45 | 6 | 18.27 | 0.03 | $\geq$0.40 | 4 | 15.83 | 0.02 | $\geq$0.22 | 5 | 13.79 | 0.02 | $\geq$0.12 | 5 |
| T9.0 | 19.26 | 0.03 | $\geq$0.88 | 4 | 18.39 | 0.03 | $\geq$0.95 | 4 | 18.77 | 0.03 | $\geq$1.08 | 4 | 18.89 | 0.08 | $\geq$0.57 | 4 | 16.17 | 0.05 | $\geq$0.23 | 2 | 14.09 | 0.04 | $\geq$0.20 | 2 |
| T9.5 | ... | ... | ... | 0 | 17.68 | 0.33 | $\leq$0.37 | 3 | 18.08 | 0.33 | $\leq$0.39 | 3 | ... | ... | ... | 0 | 15.58 | 0.37 | $\leq$0.41 | 2 | 13.51 | 0.37 | $\leq$0.43 | 2 |
| Y0.0 | 20.24 | 0.17 | $\leq$0.17 | 4 | 20.09 | 0.17 | $\leq$0.25 | 4 | 20.60 | 0.17 | $\leq$0.25 | 4 | 20.70 | 0.18 | $\leq$0.16 | 4 | 16.99 | 0.19 | $\leq$0.21 | 3 | 14.66 | 0.19 | $\leq$0.28 | 3 |

Note. — For each spectral type and bandpass four numbers are given: (1) the weighted mean absolute magnitude ("avg"); (2) error on the weighted mean ($\sigma_{avg}$); (3) the additional scatter ($\sigma_{add}$) needed to make $p(\chi^2) = 0.5$; and (4) the number of objects from Table S2 used in the bin ($n$). For cases where the rms is larger than can be explained by measurement error, i.e., $p(\chi^2) < 0.5$, the $\sigma_{add}$ values are essentially lower limits on the amount of intrinsic scatter in the absolute magnitudes at that spectral type. For cases where the rms can be explained simply by measurement uncertainties, i.e., $p(\chi^2) \geq 0.5$, the $\sigma_{add}$ values represent the upper limit on the amount of additional intrinsic scatter that could be tolerated while still keeping $p(\chi^2) \geq 0.5$. Only "normal" objects are used for these averages, i.e., excluding objects typed as peculiar (2MASS J0729−3954, BD+01 2920B, and WISEP J1405+5534) and two Y0 dwarfs with very uncertain or preliminary distances (WISE J0359−5401 and WISE J1639−6847).

Table S4. Bolometric Corrections

| Spec. Type | $YJH$ (mag) | $YJ$ (mag) | $JH$ (mag) | $Y$ (mag) | $J$ (mag) | $H$ (mag) |
|---|---|---|---|---|---|---|
| T8.0 | 2.0 | 2.0 | 2.3 | 1.6 | 2.4 | 2.1 |
| T8.5 | 1.6 | 1.6 | 1.9 | 1.1 | 2.0 | 1.6 |
| T9.0 | 1.6 | 1.5 | 1.9 | 1.0 | 2.0 | 1.7 |
| T9.5 | ⋯ | ⋯ | 1.8 | ⋯ | 1.9 | 1.5 |
| Y0.0 | 0.8 | 0.9 | 0.7 | 0.8 | 0.9 | 0.4 |
| Y0.5 | −0.7 | −0.6 | −0.7 | −0.7 | −0.4 | −1.5 |

Note. — These bolometric corrections were derived from single objects with photometry in at least $J$, $H$, [3.6], and [4.5] bands, i.e., capturing $\gtrsim$50% of the bolometric flux. We used model atmospheres (*12, 13*) to compute their bolometric magnitudes and thereby bolometric corrections in near-IR bands: $BC_X = m_{\rm bol} - m_X$. The values listed here are the weighted averages of values derived for individual objects in each spectral type bin. The rms about these weighted averages was 0.6 mag, so we take this as the uncertainty in these bolometric corrections.

Table S5. Derived Properties of Late-T and Y dwarfs with Parallaxes

| Object | Spec. Type | $m_{\rm bol}$ (mag) | $\log(L_{\rm bol}/L_\odot)$ | Cond ($t=1$ Gyr) | | | | Cond ($t=5$ Gyr) | | | |
|---|---|---|---|---|---|---|---|---|---|---|---|
| | | | | $T_{\rm eff}$ (K) | $R_\star$ ($R_{\rm Jup}$) | $M_\star$ ($M_{\rm Jup}$) | $\log g$ (cgs) | $T_{\rm eff}$ (K) | $R_\star$ ($R_{\rm Jup}$) | $M_\star$ ($M_{\rm Jup}$) | $\log g$ (cgs) |
| PSO J043.5395+02.3995 | T8 | $18.48 \pm 0.09$ | $-5.75^{+0.11}_{-0.10}$ | $660 \pm 40$ | $0.983^{+0.007}_{-0.006}$ | $17.0^{+1.7}_{-1.5}$ | $4.63 \pm 0.05$ | $720^{+50}_{-40}$ | $0.830 \pm 0.012$ | $39^{+4}_{-3}$ | $5.14 \pm 0.05$ |
| 2MASSI J0415195$-$093506 | T8 | $17.85 \pm 0.08$ | $-5.72 \pm 0.03$ | $673^{+14}_{-13}$ | $0.9820 \pm 0.0017$ | $17.4 \pm 0.5$ | $4.64 \pm 0.01$ | $734 \pm 16$ | $0.827 \pm 0.004$ | $40.1^{+1.2}_{-1.1}$ | $5.15 \pm 0.01$ |
| 2MASS J07290002$-$3954043 | T8p | $18.28 \pm 0.09$ | $-5.61 \pm 0.07$ | $720 \pm 30$ | $0.976 \pm 0.003$ | $19.2^{+1.2}_{-1.1}$ | $4.69 \pm 0.03$ | $790 \pm 30$ | $0.815^{+0.006}_{-0.007}$ | $43.7^{+2.1}_{-2.0}$ | $5.20 \pm 0.03$ |
| 2MASS J09393548$-$2448279 | T8 | $17.68 \pm 0.08$ | $-5.71 \pm 0.04$ | $678^{+17}_{-16}$ | $0.981 \pm 0.002$ | $17.6 \pm 0.6$ | $4.64^{+0.01}_{-0.01}$ | $740^{+20}_{-19}$ | $0.825 \pm 0.005$ | $40.5 \pm 1.4$ | $5.16 \pm 0.02$ |
| 2MASSW J1225543$-$273947B | T8 | $18.9 \pm 0.6$ | $-5.41 \pm 0.23$ | $810^{+130}_{-110}$ | $0.962^{+0.016}_{-0.022}$ | $23^{+5}_{-4}$ | $4.79 \pm 0.11$ | $890^{+140}_{-120}$ | $0.80 \pm 0.02$ | $50^{+8}_{-7}$ | $5.28 \pm 0.09$ |
| Ross 458C | T8 | $19.24 \pm 0.08$ | $-5.66 \pm 0.04$ | $649 \pm 14$[a] | $1.142 \pm 0.004$[a] | $6.8 \pm 0.3$[a] | $4.10 \pm 0.02$[a] | $692 \pm 16$[a] | $1.006^{+0.006}_{-0.005}$[a] | $15.9^{+0.6}_{-0.9}$[a] | $4.58^{+0.02}_{-0.03}$[a] |
| BD+01 2920B | T8p | $20.75 \pm 0.08$ | $-5.93 \pm 0.03$ | $592 \pm 14$ | $1.008 \pm 0.008$ | $13.4^{+0.9}_{-0.8}$ | $4.50 \pm 0.03$ | $644^{+15}_{-14}$ | $0.851 \pm 0.004$ | $33.8 \pm 1.0$ | $5.05 \pm 0.01$ |
| ULAS J003402.77$-$005206.7 | T8.5 | $20.35 \pm 0.08$ | $-5.91 \pm 0.04$ | $598^{+16}_{-15}$ | $1.004 \pm 0.009$ | $13.8^{+1.0}_{-0.9}$ | $4.52 \pm 0.04$ | $651^{+16}_{-15}$ | $0.849 \pm 0.004$ | $34.2 \pm 1.1$ | $5.06 \pm 0.01$ |
| CFBDS J005910.90$-$011401.3 | T8.5 | $19.80 \pm 0.08$ | $-6.04 \pm 0.04$ | $550 \pm 12$ | $1.0170 \pm 0.0005$ | $11.9 \pm 0.3$ | $4.44 \pm 0.01$ | $596 \pm 14$ | $0.865 \pm 0.005$ | $30.4^{+1.2}_{-1.1}$ | $4.99 \pm 0.02$ |
| WISEP J045853.90+643451.9A | T8.5 | $19.5 \pm 0.6$ | $-5.6^{+0.4}_{-0.3}$ | $730^{+180}_{-130}$ | $0.98 \pm 0.03$ | $20^{+8}_{-6}$ | $4.70^{+0.17}_{-0.18}$ | $800^{+200}_{-150}$ | $0.81^{+0.04}_{-0.03}$ | $44^{+12}_{-10}$ | $5.21^{+0.14}_{-0.15}$ |
| WISE J111838.70+312537.9 | T8.5 | $19.56 \pm 0.08$ | $-6.08 \pm 0.04$ | $538 \pm 11$ | $1.0176 \pm 0.0005$ | $11.6 \pm 0.3$ | $4.43 \pm 0.01$ | $582 \pm 14$ | $0.870 \pm 0.005$ | $29.2 \pm 1.1$ | $4.97 \pm 0.02$ |
| ULAS J133553.45+113005.2 | T8.5 | $19.93 \pm 0.08$ | $-6.07 \pm 0.04$ | $542 \pm 11$ | $1.0174 \pm 0.0005$ | $11.7 \pm 0.3$ | $4.43 \pm 0.01$ | $587^{+14}_{-13}$ | $0.869 \pm 0.005$ | $29.6 \pm 1.1$ | $4.97 \pm 0.02$ |
| Wolf 940B | T8.5 | $20.39 \pm 0.08$ | $-6.10 \pm 0.05$ | $532^{+17}_{-16}$ | $1.0178^{+0.0007}_{-0.0008}$ | $11.5 \pm 0.4$ | $4.42 \pm 0.01$ | $575^{+20}_{-19}$ | $0.873 \pm 0.007$ | $28.6^{+1.6}_{-1.5}$ | $4.96 \pm 0.03$ |
| UGPS J072227.51$-$054031.2 | T9 | $18.33 \pm 0.08$ | $-6.20 \pm 0.03$ | $502 \pm 10$ | $1.0192 \pm 0.0005$ | $10.7 \pm 0.2$ | $4.39 \pm 0.01$ | $539 \pm 12$ | $0.886 \pm 0.005$ | $25.8 \pm 0.9$ | $4.90 \pm 0.01$ |
| WISEP J121756.91+162640.2A | T9 | $19.9 \pm 0.6$ | $-6.04^{+0.29}_{-0.28}$ | $550^{+110}_{-80}$ | $1.017^{+0.011}_{-0.034}$ | $12^{+5}_{-2}$ | $4.44^{+0.19}_{-0.11}$ | $600^{+120}_{-100}$ | $0.86 \pm 0.04$ | $31^{+9}_{-8}$ | $4.99^{+0.15}_{-0.16}$ |
| CFBDSIR J145829+101343A | T9 | $21.8 \pm 0.6$ | $-5.82 \pm 0.26$ | $630^{+110}_{-100}$ | $0.987^{+0.030}_{-0.013}$ | $16 \pm 4$ | $4.60^{+0.12}_{-0.17}$ | $690^{+120}_{-110}$ | $0.84 \pm 0.04$ | $37 \pm 8$ | $5.10^{+0.12}_{-0.13}$ |
| WISEP J174124.26+255319.5 | T9 | $18.30 \pm 0.08$ | $-5.93 \pm 0.08$ | $590 \pm 30$ | $1.009^{+0.008}_{-0.019}$ | $13.3^{+2.2}_{-1.1}$ | $4.50^{+0.08}_{-0.04}$ | $640^{+40}_{-30}$ | $0.851^{+0.009}_{-0.010}$ | $34 \pm 2$ | $5.05 \pm 0.04$ |
| WISEP J014807.25$-$720258.8 | T9.5 | $20.79 \pm 0.08$ | $-5.97^{+0.27}_{-0.21}$ | $570^{+110}_{-70}$ | $1.016^{+0.003}_{-0.035}$ | $12.5^{+5.3}_{-1.6}$ | $4.47^{+0.19}_{-0.07}$ | $630^{+120}_{-80}$ | $0.86 \pm 0.03$ | $33^{+8}_{-6}$ | $5.03^{+0.14}_{-0.11}$ |
| WISEP J045853.90+643451.9B | T9.5 | $20.4 \pm 0.6$ | $-5.9^{+0.4}_{-0.3}$ | $590^{+150}_{-100}$ | $1.009^{+0.011}_{-0.035}$ | $13^{+7}_{-3}$ | $4.50^{+0.21}_{-0.12}$ | $640^{+170}_{-120}$ | $0.85 \pm 0.04$ | $34^{+11}_{-9}$ | $5.05 \pm 0.17$ |
| WISEP J180435.40+311706.1 | T9.5: | $20.67 \pm 0.08$ | $-5.92^{+0.18}_{-0.15}$ | $590^{+70}_{-50}$ | $1.007^{+0.010}_{-0.024}$ | $13.5^{+3.6}_{-1.8}$ | $4.51^{+0.13}_{-0.07}$ | $650^{+80}_{-60}$ | $0.850^{+0.019}_{-0.021}$ | $34^{+6}_{-4}$ | $5.05^{+0.09}_{-0.08}$ |
| WISE J035934.06$-$540154.6 | Y0 | $21.79 \pm 0.08$ | $-7.27^{+0.18}_{-0.13}$ | $265^{+28}_{-19}$ | $1.065 \pm 0.005$ | $3.0^{+0.8}_{-0.5}$ | $3.81^{+0.10}_{-0.07}$ | $270^{+30}_{-20}$ | $0.992^{+0.008}_{-0.009}$ | $7.9^{+1.8}_{-1.1}$ | $4.29^{+0.10}_{-0.07}$ |
| WISEP J041022.71+150248.5 | Y0 | $20.53 \pm 0.08$ | $-6.55^{+0.11}_{-0.10}$ | $400 \pm 30$ | $1.047^{+0.018}_{-0.004}$ | $7.4^{+0.8}_{-1.0}$ | $4.21^{+0.05}_{-0.08}$ | $430 \pm 30$ | $0.929^{+0.010}_{-0.011}$ | $17.8^{+2.2}_{-1.8}$ | $4.70 \pm 0.06$ |
| WISEP J121756.91+162640.2B | Y0 | $21.0 \pm 0.6$ | $-6.49^{+0.29}_{-0.28}$ | $420^{+80}_{-70}$ | $1.04^{+0.02}_{-0.03}$ | $8^{+3}_{-2}$ | $4.24^{+0.15}_{-0.16}$ | $450^{+90}_{-70}$ | $0.92 \pm 0.04$ | $19^{+7}_{-5}$ | $4.73^{+0.17}_{-0.18}$ |
| WISEP J140518.40+553421.5 | Y0p | $20.61 \pm 0.08$ | $-6.56^{+0.14}_{-0.12}$ | $400^{+40}_{-30}$ | $1.048^{+0.019}_{-0.006}$ | $7.3^{+1.1}_{-1.2}$ | $4.21^{+0.06}_{-0.09}$ | $430^{+40}_{-30}$ | $0.930 \pm 0.014$ | $18^{+3}_{-2}$ | $4.70 \pm 0.08$ |
| WISE J163940.83$-$684738.6 | Y0: | $21.7 \pm 0.7$ | $-7.38^{+0.28}_{-0.27}$ | $250 \pm 40$ | $1.061^{+0.009}_{-0.013}$ | $2.6^{+1.2}_{-0.7}$ | $3.76^{+0.14}_{-0.15}$ | $260^{+50}_{-40}$ | $0.999^{+0.014}_{-0.016}$ | $7.0^{+2.7}_{-1.8}$ | $4.23^{+0.16}_{-0.13}$ |
| WISEP J173835.52+273258.9 | Y0 | $20.87 \pm 0.08$ | $-6.46^{+0.17}_{-0.14}$ | $430^{+50}_{-40}$ | $1.043^{+0.013}_{-0.018}$ | $8.1^{+1.8}_{-1.2}$ | $4.25^{+0.10}_{-0.08}$ | $450^{+50}_{-40}$ | $0.920^{+0.015}_{-0.021}$ | $20^{+4}_{-3}$ | $4.75^{+0.10}_{-0.09}$ |
| WISEP J205628.90+145953.3 | Y0 | $20.25 \pm 0.08$ | $-6.52^{+0.15}_{-0.13}$ | $410^{+40}_{-30}$ | $1.046^{+0.015}_{-0.012}$ | $7.6^{+1.4}_{-1.3}$ | $4.23^{+0.08}_{-0.09}$ | $440^{+50}_{-40}$ | $0.925^{+0.013}_{-0.017}$ | $19^{+3}_{-2}$ | $4.72^{+0.09}_{-0.08}$ |
| WISEP J154151.65$-$225025.2 | Y0.5 | $20.72 \pm 0.08$ | $-6.1^{+0.5}_{-0.3}$ | $530^{+180}_{-90}$ | $1.02^{+0.02}_{-0.04}$ | $11^{+7}_{-3}$ | $4.42^{+0.26}_{-0.15}$ | $570^{+200}_{-100}$ | $0.88^{+0.04}_{-0.06}$ | $28^{+15}_{-8}$ | $4.95^{+0.24}_{-0.17}$ |
| WISE J035000.32$-$565830.2 | Y1 | $21.05 \pm 0.21$ | $-7.38^{+0.32}_{-0.13}$ | $249^{+49}_{-17}$ | $1.060^{+0.010}_{-0.005}$ | $2.6^{+1.4}_{-0.4}$ | $3.75^{+0.18}_{-0.07}$ | $257^{+55}_{-19}$ | $1.000^{+0.005}_{-0.019}$ | $6.9^{+3.1}_{-0.9}$ | $4.22^{+0.18}_{-0.06}$ |
| WISE J053516.80$-$750024.9 | $\geq$Y1: | $21.32 \pm 0.21$ | $-6.0^{+0.4}_{-0.6}$ | $570^{+170}_{-190}$ | $1.02 \pm 0.04$ | $12^{+8}_{-6}$ | $4.5 \pm 0.3$ | $620^{+200}_{-210}$ | $0.86^{+0.08}_{-0.05}$ | $32^{+13}_{-16}$ | $5.0^{+0.2}_{-0.4}$ |
| WISEP J182831.08+265037.8 | $\geq$Y2 | $20.86 \pm 0.08$ | $-6.13^{+0.20}_{-0.16}$ | $520^{+60}_{-50}$ | $1.018^{+0.007}_{-0.006}$ | $11.2^{+1.9}_{-1.4}$ | $4.41^{+0.08}_{-0.06}$ | $560^{+80}_{-60}$ | $0.88^{+0.02}_{-0.03}$ | $28^{+6}_{-4}$ | $4.94^{+0.11}_{-0.09}$ |
| CFBDSIR J145829+101343B | $\cdots$ | $23.0 \pm 0.6$ | $-6.27 \pm 0.26$ | $480^{+80}_{-70}$ | $1.023^{+0.023}_{-0.006}$ | $10^{+2}_{-3}$ | $4.36^{+0.09}_{-0.14}$ | $520^{+100}_{-80}$ | $0.90^{+0.03}_{-0.04}$ | $24^{+7}_{-6}$ | $4.86 \pm 0.15$ |
| WD 0806$-$661B | $\cdots$ | $23.19 \pm 0.21$ | $-6.81 \pm 0.09$ | $350^{+20}_{-19}$[b] | $1.040^{+0.009}_{-0.008}$[b] | $6.8 \pm 0.7$[b] | $4.19 \pm 0.05$[b] | $356^{+20}_{-19}$[b] | $1.007^{+0.007}_{-0.009}$[b] | $9.0^{+1.0}_{-0.9}$[b] | $4.33 \pm 0.05$[b] |

Note. — Uncertainties in spectral types are 0.5 subtypes unless otherwise noted ($\pm 1$ subtype errors are denoted by ":"). Apparent bolometric magnitudes ($m_{\rm bol}$) and bolometric luminosities are computed as described in the text. Effective temperatures ($T_{\rm eff}$), radii ($R_\star$), masses ($M_\star$), and surface gravities ($\log g$) are interpolated from Cond model isochrones (14) at ages of 1 Gyr and 5 Gyr. Error bars on these model derived properties only reflect the nominal uncertainty in the luminosity, i.e., due to the rms in parallax and bolometric corrections, at the given age. They do not include any potential systematic errors in bolometric corrections or the evolutionary model isochrones. Note that we assumed a spectral type of Y0 for CFBDSIR J145829+101343B here, solely for the purpose of computing $m_{\rm bol}$.

[a] For Ross 458C ages of 150 Myr and 800 Myr were used instead of 1 Gyr and 5 Gyr, based on the age range determined by (21).

[a] For WD 0806$-$661B ages of 1.5 Gyr and 2.5 Gyr were used instead of 1 Gyr and 5 Gyr, based on the age range determined by (22).

Table S6. Mean Properties of Normal Late-T and Y dwarfs

| Spec. Type | $\log(L_{bol}/L_\odot)$ | Cond ($t = 1$ Gyr) | | | | | Cond ($t = 5$ Gyr) | | | | |
|---|---|---|---|---|---|---|---|---|---|---|---|
| | | $T_{eff}$ (K) | $R_\star$ ($R_{Jup}$) | $M_\star$ ($M_{Jup}$) | $\log g$ (cgs) | $D/D_0$ | $T_{eff}$ (K) | $R_\star$ ($R_{Jup}$) | $M_\star$ ($M_{Jup}$) | $\log g$ (cgs) | $D/D_0$ |
| T8.0 | $-5.72$ | 675 | 0.982 | 17 | 4.65 | 0.00 | 735 | 0.826 | 40 | 5.16 | 0.00 |
| T8.5 | $-6.04$ | 550 | 1.017 | 12 | 4.44 | 0.29 | 600 | 0.864 | 31 | 5.00 | 0.00 |
| T9.0 | $-6.15$ | 510 | 1.019 | 11 | 4.40 | 0.76 | 555 | 0.879 | 27 | 4.94 | 0.00 |
| T9.5 | $-5.94$ | 585 | 1.011 | 13 | 4.49 | 0.27 | 640 | 0.852 | 33 | 5.05 | 0.00 |
| Y0.0 | $-6.52$ | 410 | 1.046 | 8 | 4.23 | 1.00 | 440 | 0.926 | 18 | 4.72 | 0.00 |

Note. — Only "normal" objects are used for these weighted averages, i.e., excluding objects typed as peculiar (2MASS J0729−3954, BD+01 2920B, and WISEP J1405+5534), two Y0 dwarfs with very uncertain or preliminary distances (WISE J0359−5401 and WISE J1639−6847), and the young T8 dwarf Ross 458C. Averages are computed from the model-derived properties listed in Table S5, except for the deuterium abundance relative to the initial abundance (D/D$_0$), which is only reported here. Effective temperatures are rounded to the nearest 5 K.




\end{document}